\documentclass[ESD,manuscript]{copernicus}


\usepackage{booktabs}
\usepackage{graphicx}
\usepackage{xr}
\usepackage{cleveref}
\usepackage{xr}

\runningtitle{TEXT}
\runningauthor{TEXT}

\received{}
\pubdiscuss{} %% only important for two-stage journals
\revised{}
\accepted{}
\published{}

\firstpage{1}
\usepackage{booktabs}
\begin{document}
\nolinenumbers
%\title{Multiple climate attractors in the present-day continental configuration: mechanisms and implications}
%\title{Five climates, one Earth}

\title{Competing climates of the present-day Earth: from Snowball to Hothouse}

%\Author[affil]{given_name}{surname}

\Author[1,2][laure.moinat@unige.ch]{Laure}{Moinat} %% correspondence author
\Author[3]{John}{Marshall}
\Author[1,2]{Maura}{Brunetti}

\affil[1]{Group of Applied Physics and Institute for Environmental Sciences, University of Geneva, Bd. Carl-Vogt 66, 1205 Geneva, Switzerland}
\affil[2]{Centre pour la Vie dans l'Univers (CVU), University of Geneva, Geneva, Switzerland}
\affil[3]{Department of Earth, Atmospheric, and Planetary Sciences, Massachusetts Institute of Technology, Cambridge, Massachusetts, USA}

%% The [] brackets identify the author with the corresponding affiliation. 1, 2, 3, etc. should be inserted.

%% If an author is deceased, please add \deceased[$Deceased date if applicable$]{$Author number$} (e.g. \deceased[13 November 2015]{2}) at the end of the affiliations. The author number depends on the placement of the author in the author list, e.g. the third author has number 3.

%% If authors contributed equally, please add \equalcontrib{$Author numbers$} (e.g. \equalcontrib{1,3}) at the end of the affiliations. The author number depends on the placement of the author in the author list, e.g. the third author has number 3.

\runningtitle{TEXT}
\runningauthor{TEXT}

\received{}
\pubdiscuss{} %% only important for two-stage journals
\revised{}
\accepted{}
\published{}

%% These dates will be inserted by Copernicus Publications during the typesetting process.

\firstpage{1}

\maketitle

\begin{abstract}
%%%%%
Earth today has one climate, but it could have others. With the same continents, the same solar constant and the same pre-industrial CO$_2$, our planet could settle into strikingly different ones. We explore this possibility with BIG-MITgcm, a model which couples atmosphere and ocean circulation with ice sheets and vegetation, and which can be integrated over many millennia. We find five different climates and describe them here. At one extreme is a Snowball, with ice from pole to pole. A Waterbelt also exists, in which sea ice reaches the tropics and only a thin equatorial ribbon of open ocean survives. There is a Cold world much like the Last Glacial Maximum, a Warm climate in which we live in today, and a Hot world in which the Atlantic overturning circulation has collapsed. Each state is held in place by its own balance of competing feedbacks. Ocean heat transport anchors each climate, holding the ice edge where that transport is weak. The familiar climate of the past 10 thousand years is perhaps just one of several possibilities. This has implications for glacial-interglacial transitions and the risk of crossing global tipping points in the future.
\end{abstract}

%\copyrightstatement{TEXT} %% This section is optional and can be used for copyright transfers.

\introduction %% \introduction[modified heading if necessary]
Since the seminal papers of \citet{Budyko_1969} and \citet{Seller_1969}, the climate system has been shown to exhibit regions of multi-stability, in which different initial conditions can result in distinct steady states under the same forcing. 
In the original Budyko and Sellers Energy Balance Model (EBM), diffusive (or similar) meridional heat transport, combined with ice-albedo and Planck feedbacks, gives rise to multiple equilibria: a warm ice-free state, a cold Snowball state, and an intermediate “large ice cap” with ice extending into mid-latitudes that is unstable—any small equatorward perturbation of the ice edge triggers rapid global glaciation due to the runaway albedo effect. \citet{Rose_Marshall_2009} extended the EBM by separating atmospheric and oceanic heat transports, explicitly including sea ice as an insulating layer between the ocean and the atmosphere, and prescribing a meridional ocean heat transport profile with strong poleward transport from the tropics and convergence at mid-latitudes (associated with wind-driven gyres, such as at the subtropical-subpolar boundary). This mid-latitude convergence of ocean heat transport supplies sufficient heat to the ice edge to counteract the albedo feedback, stabilizing the large ice cap and creating a stable equilibrium where the ice margin rests at a preferred latitude (near the zero wind curl line) without triggering runaway ice expansion. 

EBMs are highly simplified and cannot capture the richness of the myriad possible interactions between climate components. The question then arises as to whether multiple equilibria exist in more complex models. Today, there is a hierarchy of models of varying levels of complexity ranging from simple to those capable of representing many nonlinear processes and feedback mechanisms. As a result, the same forcing may lead to different balances among these feedbacks, allowing the climate system to relax toward distinct steady states. Indeed, multi-stability has been previously observed in Earth system models of intermediate complexity (EMICs --- see \citet{rahmstorf2005,Lucarini_2017}), as well as in general circulation models (GCMs) under aquaplanet or simplified continental configurations --- \citep{Ferreira2011, Rose2015, Brunetti_2019, Ragon_2022}. These include setups with limited land strips, such as the `Ridge’ configuration (a meridional land strip) or the `Drake’ configuration (a meridional land strip extending from the North Pole to 60$^\circ$S). In the aquaplanet configuration, up to five distinct steady states have been identified, including snowball, waterbelt, cold, warm, and hot states, with the large ice cap stabilized by the meridional structure of ocean heat transport convergence. The existence of these multiple equilibria is robust in different regimes of ocean circulation~\citep{Ferreira2011, Brunetti_2019}. Subsequent studies have built on these insights to explore the role of ocean and atmosphere heat transport in sea-ice sensitivity, waterbelt climates, and paleoclimate transitions~\citep{Rose2015,zhurose2023,RagonPT,horner2025}.

The existence of steady states can act as attractors for climate trajectories evolving in different regions of phase space~\citep{Milnor1985,Gallavotti2006}. Each attractor corresponds to a climate regime associated with a different degree of internal variability and embedded in a high-dimensional phase space defined by relevant climatic variables, such as temperature, precipitation, and ice extent. The basin of attraction associated with each attractor partitions the phase space into stable and unstable regions. The latter are particularly relevant for understanding how and why the climate system may shift from one state to another in response to internal variability, or small perturbations of forcing intensity or rate (so-called noise-induced, bifurcation-induced and rate-induced tipping points, respectively)~\citep{ashwin2012,Ragon_2023}. For this reason, identifying the dynamical structure of the climate system, namely its possible attractors, their basins of attraction, and the boundaries between them, is crucial for understanding climate evolution and the conditions that can drive it towards stable or unstable trajectories. The extent of the basins of attraction and their boundaries depend on the continental configuration and on the number, strength, and representation of feedback processes. 

Here, we investigate the dynamical structure of the climate system under the present-day continental configuration and focus on whether the pre-industrial climate regime is the only possible realization under an atmospheric  CO$_2$ content of around 280~ppm over a multi-millennial timescale. The present continental configuration has been approximately constant over the past million years \citep{Muller2016,VERARD_2019,paleomap,reconstruction_meredith}, the plate displacements during that interval being only \(\mathcal{O}(10^{1}\text{--}10^{2})\) km. Previous studies have explored the dynamical structure of the climate system using EMICs but without accounting for vegetation and ice sheets. Here, we employ a new coupled framework, called {\it BIG-MITgcm} -- see~\citet{Moinat_2026} and Section~\ref{sec:model} -- with representation of vegetation and ice sheets, but also atmosphere-ocean dynamics, the hydrological cycle, sea ice, and biogeochemical cycles. We find that the pre-industrial state is indeed not the only climate regime; additional four steady states under the same radiative forcing (220-280 ppm) are possible, ranging from a snowball state~\citep{Snowball_Pierrehumbert}, where the Earth is completely covered by ice sheets and sea ice, a waterbelt with an equatorial ocean, a cold state resembling glacial conditions, and a hot state representing a potential future climate in which the Atlantic Meridional Overturning Circulation (AMOC) has collapsed. We describe the equilibrium states and in particular how the land ice and sea ice distributions vary across them, along with the associated rearrangements of coupled dynamics and the carbon cycle. We also discuss mechanisms that allow multiple states to be sustained under similar external forcing and the implications of their existence for past and future climate trajectories. Particular emphasis is placed on the role of ocean heat transport in stabilizing the ice edge in each of our equilibrium solutions.

%%%%%%%%%%%%%%%%%%%%%%%%%%
\section{Nature of the equilibrium solutions}

\begin{figure}[ht!]
\includegraphics[width=\textwidth]{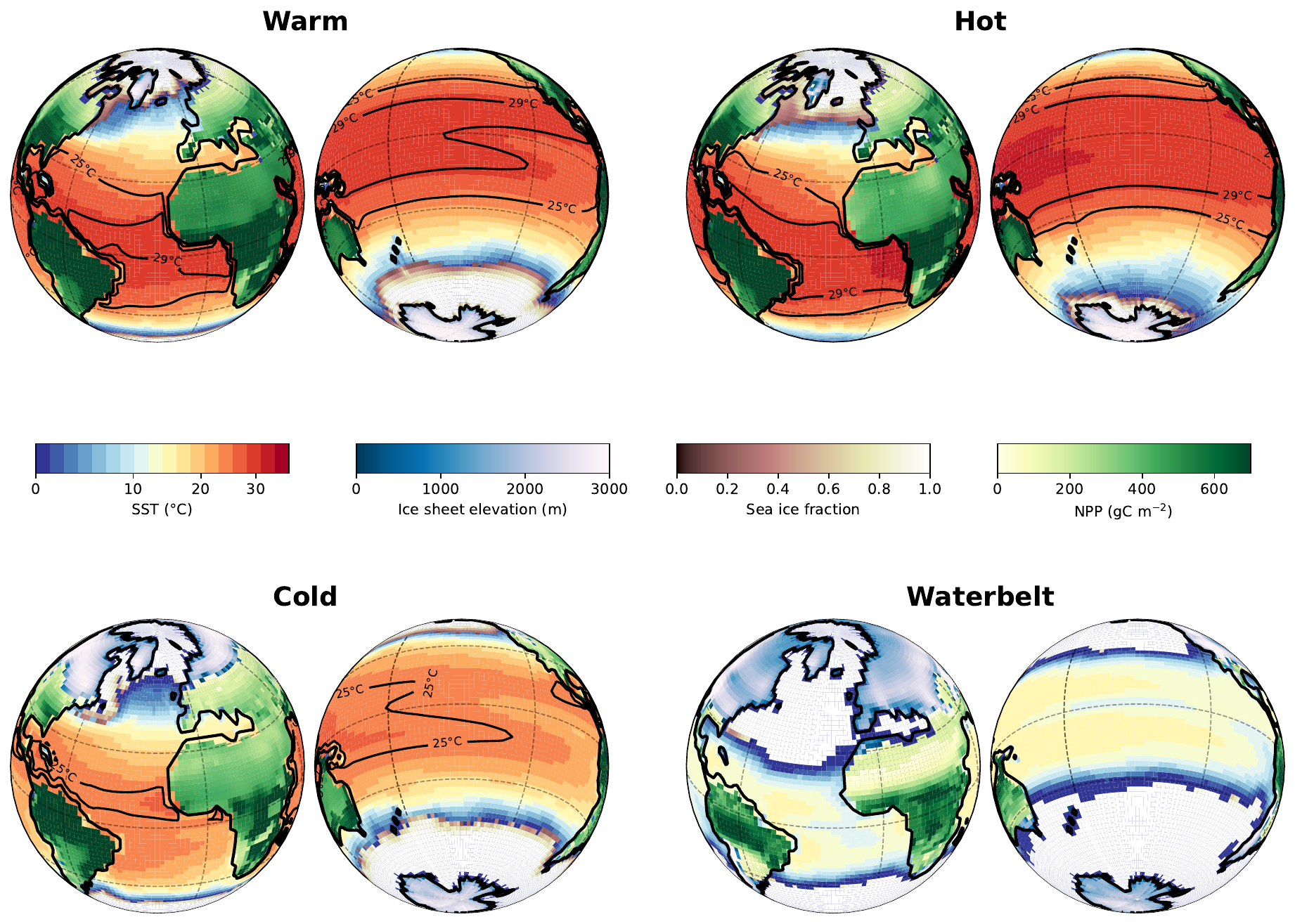}
\caption{Sea surface temperature (blue-red shading in °C), ice-sheet elevation (blue-white shading in meters), sea-ice fraction (brown-white shading) and net primary productivity (yellow-green shading in gCm$^{-2}$) distributions for the Hot (top left), Warm (top right), Cold (bottom left) and Waterbelt (bottom right) states. The sea-ice fraction has a threshold at 0.15. All states were obtained using the same parameters and external forcing, except for Cold state, which was obtained with a CO$_2$ concentration of 220 ppm.}
\label{fig:FOUR_PLANETS_ORTHO}
\end{figure}

We now describe each of the climatic steady states that exist at the same CO$_2$ concentration of 280~ppm: they are Warm, Hot, Waterbelt, and Snowball. These states coexist as multiple attractors (multistable), i.e. they can be reached from different initial conditions within phase space under the same forcing. In contrast, we find that the stable branch of the Cold state exists only at slightly lower atmospheric CO$_2$ concentrations of less than 220~ppm. We do not describe the stable branches of the attractors here; instead, we concentrate on their climatic characteristics at the chosen CO$_2$ concentrations. The existence of a multistable region of phase space implies that nonlinear feedback mechanisms differ between attractors. We include a detailed description of the Cold state at 220~ppm since it is likely relevant for glacial periods. 
These stable states are presented 
in Fig.~\ref{fig:FOUR_PLANETS_ORTHO}: the distribution of surface air temperature, ice-sheet elevation, net primary production, and sea-ice extent for Hot, Warm and Waterbelt at 280~ppm is shown, along with the cold state at 220~ppm. The Snowball state is shown in the Supplementary Section~\ref{sec:snowball}.

%%%%%%%%%%%%%%%%%%%%%%%
\subsection{Warm state}

The Warm state, which we can call \textit{piControl}, has already been compared with observations, reanalysis, and CMIP-class models in \citet{Moinat_2026}. However, because a substantial number of new diagnostics are employed here and this is the solution against which the others are to be compared, we briefly describe it. The Warm state exhibits many similarities with the present-day climate, including broadly similar tropical and polar temperatures. The tropical sea surface temperatures (SSTs) in the warm pool of the Pacific reach approximately 29°C or so and are higher than those of the tropical Atlantic (Fig.~\ref{fig:FOUR_PLANETS_ORTHO}). The zonal asymmetries in the SST within the basins reflect the influence of the ocean gyres and their western intensification. Both poles are cold and covered by sea ice. There is substantial Arctic sea ice cover and significant Antarctic sea ice that extends northward into the Drake Passage. Continental ice sheets are present over Greenland and Antarctica, somewhat corresponding to modern conditions.

\begin{figure}[ht!]
\includegraphics[width=\textwidth]{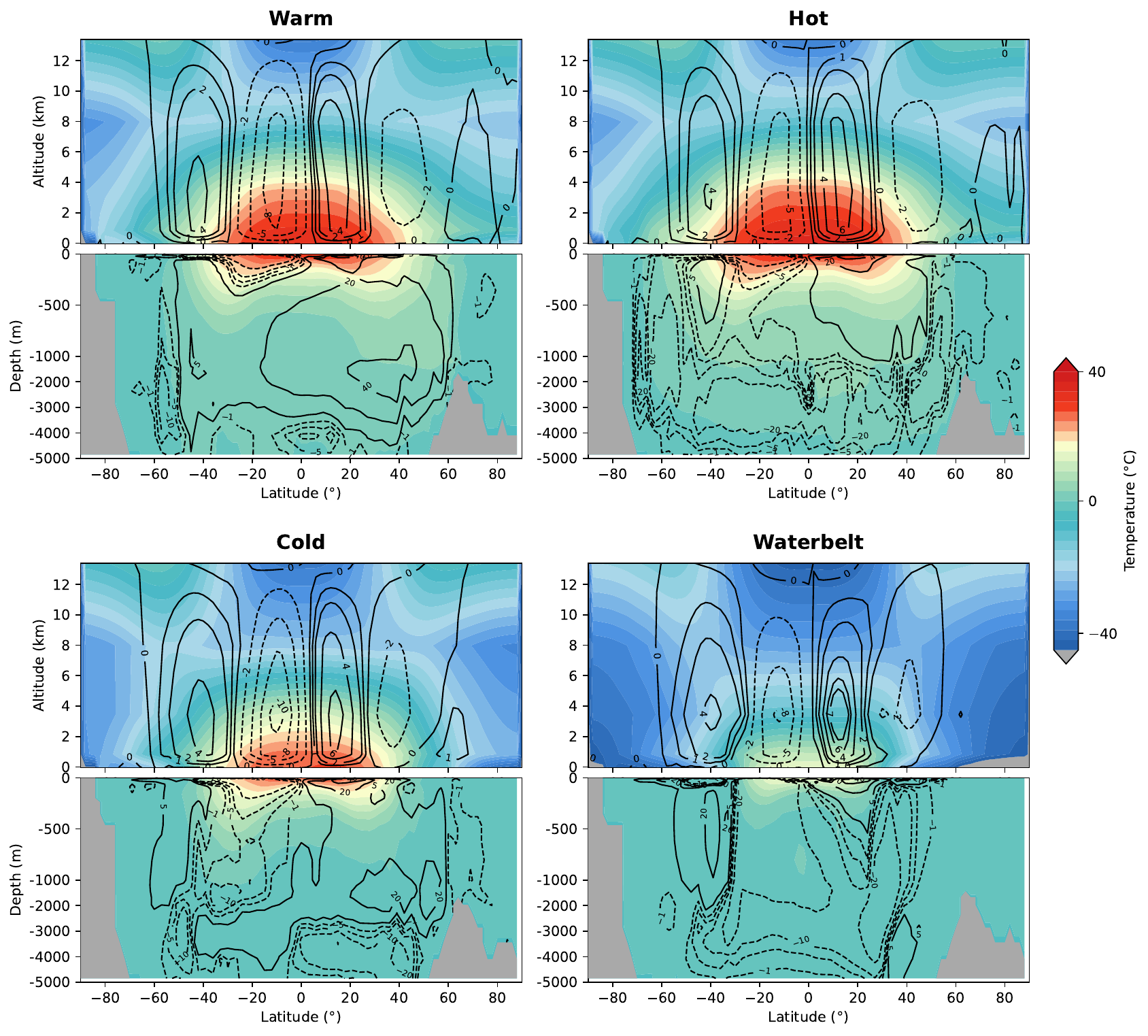}
\caption{Atmosphere and ocean temperature distributions (color shading) and global atmosphere and ocean overturning circulations for each state. The Eulerian overturning is plotted in the atmosphere and the residual in the ocean. Atmospheric and oceanic potential temperature are indicated by the blue-red shading in °C: the black contours plot the overturning cells, where the solid lines indicate positive (clockwise) cells and the dashed lines indicate negative (counter-clockwise) cells for the Warm (top left), Hot (top right), Cold (bottom left) and Waterbelt (bottom right) states, receptively.}
\label{fig:Figure3}
\end{figure}

\begin{figure}[ht!]
\includegraphics[width=\textwidth]{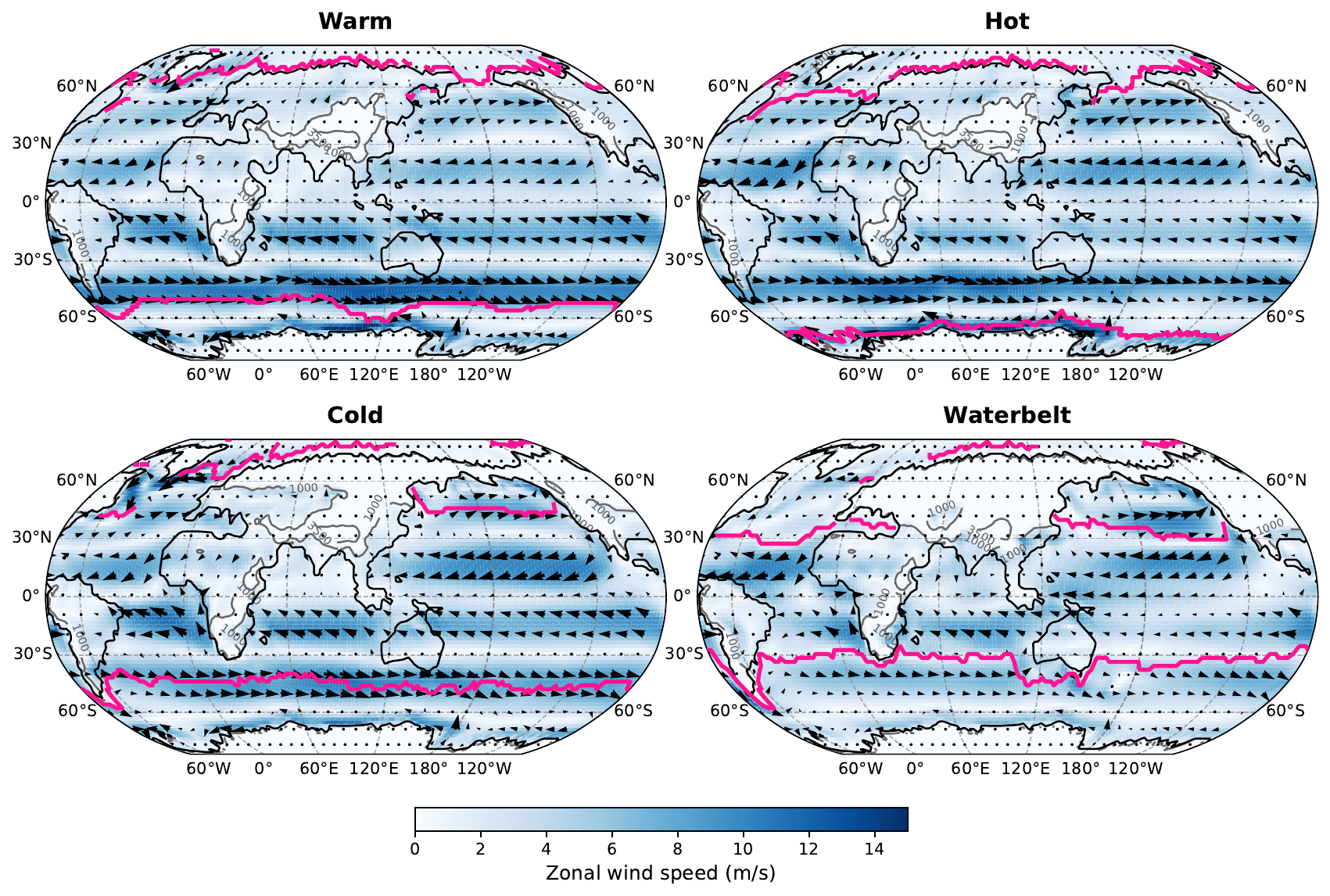}
\caption{Surface winds are shown by the blue–white shading, with arrows indicating their direction and magnitude, for the Warm, Hot, Cold, and Waterbelt states. The solid pink line represents the annual-mean sea-ice edge, while the solid grey lines indicate the 1000 m and 3500 m topographic contours, combining mountain and glacial elevations.}
\label{fig:GLOBAL_WINDS}
\end{figure}

Fig.~\ref{fig:Figure3} shows zonal-mean temperature and overturning circulation in the atmosphere and ocean for our four solutions. We see that zonal-mean atmosphere and ocean temperatures are asymmetric about the equator in the Warm state, with the Northern Hemisphere warmer than the Southern Hemisphere and the Inter-Tropical Convergence Zone (ITCZ) displaced north of the equator, consistent with key features of the present climate. The Southern Hemisphere Hadley cell is stronger than the Northern Hemisphere one, producing an asymmetric surface wind pattern consistent with the present climate: surface westerlies at mid-latitudes and easterly trade winds in the tropics, with stronger trade winds in the Southern Hemisphere (Fig.~\ref{fig:GLOBAL_WINDS}). Subpolar cyclonic surface-wind patterns are associated with low-pressure regions over the ocean, while  anticyclonic circulation develops around the surface high-pressure systems associated with ice sheets and their cold air masses (Fig.~\ref{fig:GLOBAL_WINDS}).

%%%%%%%%%%%%%%%%%%%%

Air temperatures over Antarctica are substantially colder than over the Arctic due to the cooling effect of the Antarctic Ice Sheet (Fig.~\ref{fig:Figure3}). Upper-ocean temperatures in the Northern Hemisphere are significantly warmer than their counterparts in the Southern Hemisphere, reflecting the impact of a global meridional overturning circulation (GMOC) in the ocean that transports heat across the equator from south to north (Figs.~\ref{fig:FOUR_PLANETS_ORTHO} and \ref{fig:Figure3}).

\begin{figure}[t!]
\includegraphics[width=1.0\textwidth]{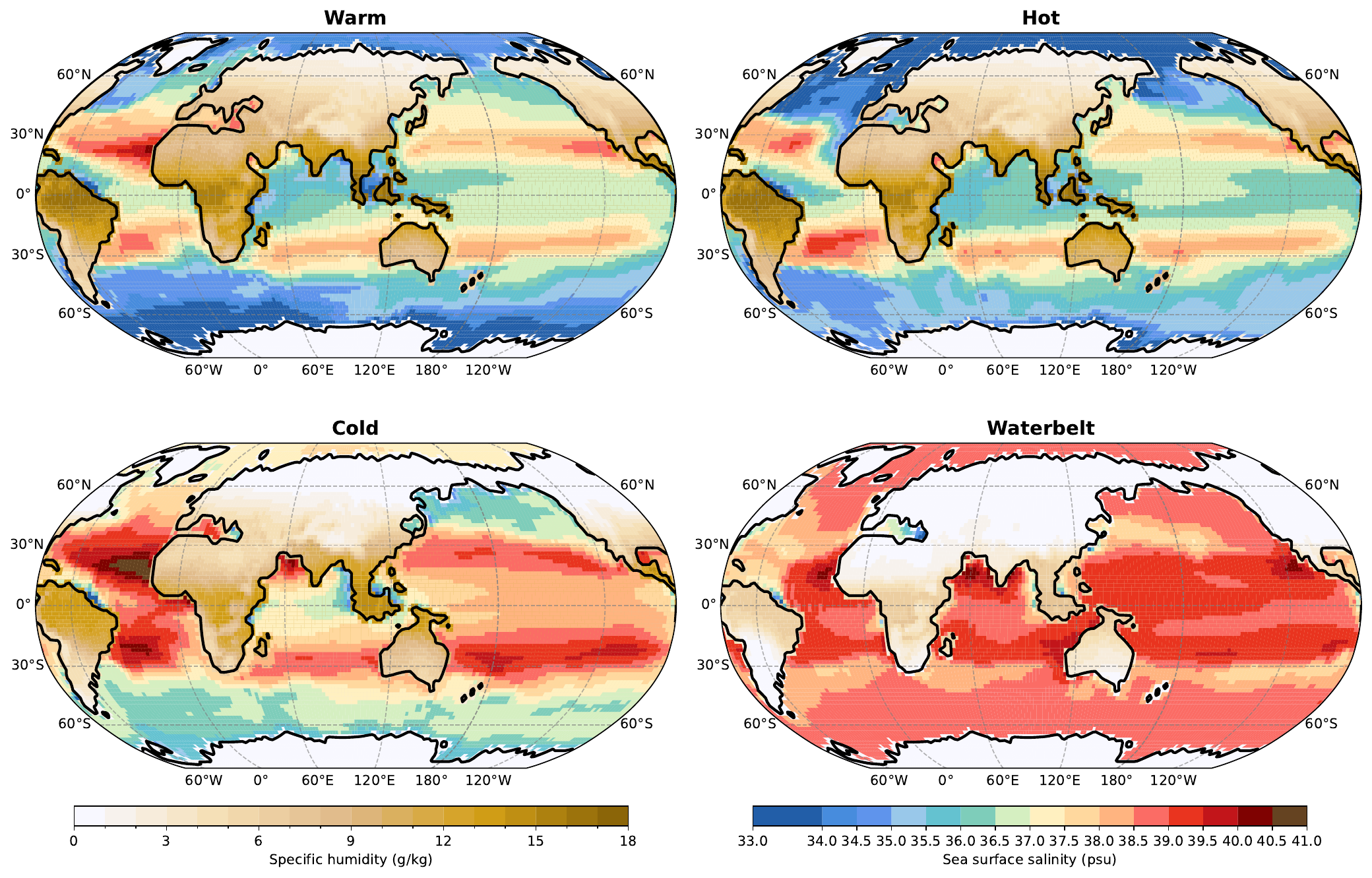}
\caption{Surface specific humidity and sea surface salinity distributions for the Warm (top left), Hot (top right), Cold (bottom left) and Waterbelt (bottom right) states. Surface specific humidity is shown by the white-gold shading in g/kg over land and sea surface salinity is shown by the blue-green-red shading in psu over the ocean. Coast lines and ocean passages are consistent with the sea level in each attractor, as evident in particular in the Bering Strait.}
\label{fig:sss_humidity}
\end{figure}

The sea surface salinity (SSS) distribution reveals bands of enhanced salinity, particularly evident in the Atlantic sector, a notable feature of our current climate (Fig.~\ref{fig:sss_humidity}). Here, strong evaporation under trade winds increases SSS (Figs.~\ref{fig:GLOBAL_WINDS} and \ref{fig:sss_humidity}). Precipitation at higher latitudes and in the deep tropics leads to surface freshening. This freshening is enhanced by sea-ice melt, particularly around Antarctica. In general, the Atlantic is markedly saltier than the Pacific, favoring the formation and sinking of dense polar waters in the Atlantic sector.

\begin{figure}[t!]
\includegraphics[width=\textwidth]{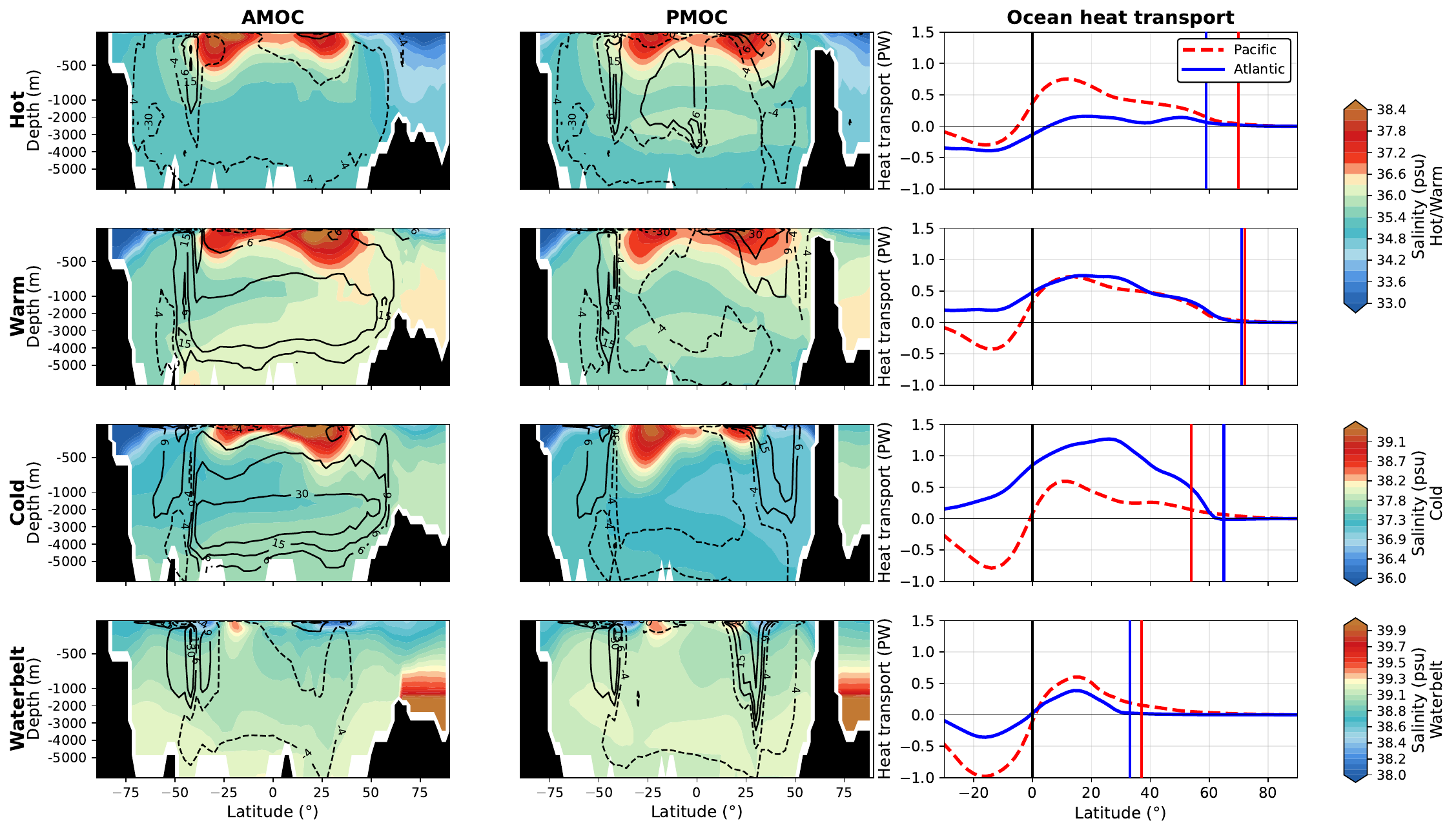}
\caption{Atlantic (first column) and Pacific (second column) Meridional
Overturning Circulations, and salinity distributions for the Hot
(first row), Warm (second row), Cold (third row) and Waterbelt
(fourth row) states. The meridional overturning circulation is
represented by black contours, with solid contours indicating positive
(clockwise) circulation and dashed contours indicating negative
(counter-clockwise) circulation. Salinity is shown in psu using
blue--green--red shading. The Hot and Warm states share the same colour
scale, while the Cold and Waterbelt states each have their own. The third column shows ocean heat transport for the Atlantic (solid blue line) and Pacific (dashed red line) basins, in PW. Vertical lines denote the sea ice edges in the respective basins in the Northern Hemisphere. }
\label{fig:AMOC-PMOC}
\end{figure}

The GMOC has an upper clockwise cell fed primarily by North Atlantic waters and a lower anticlockwise cell originating from the Southern Ocean around Antarctica. GMOC (Fig.~\ref{fig:Figure3}) and Atlantic Meridional Overturning Circulation (AMOC) (Fig.~\ref{fig:AMOC-PMOC}) exhibit patterns consistent with those found in other models and ocean reanalysis. The Atlantic exhibits a strong positive AMOC, with clear signatures in temperature and salinity of North Atlantic Deep Water (NADW) and Antarctic Intermediate Water (AAIW). The Pacific (PMOC) has a weak large-scale negative cell fed from the south, together with a shallow, wind-driven positive cell near 40°N (Fig.~\ref{fig:AMOC-PMOC}). There is also evidence of Antarctic Bottom Water (AABW) penetrating the abyss from the south; this water is colder and fresher than the overlying waters.

The ocean oxygen distribution in Fig.~\ref{fig:oxygen} reveals high surface values at the poles (due to temperature-dependent solubility of oxygen in water), depleted tropical concentrations in the upper kilometer, and deep oxygenation from both poles, particularly from the south around Antarctica in the Pacific basin and from the North in the Atlantic basin, and broadly consistent with the present-day climate. The net air-sea heat flux (Fig.~\ref{fig:Figure9}) is directed into the ocean in the tropics and out of the ocean at mid-latitudes, particularly along the western boundary and polar regions. This distribution also clearly highlights the boundary between open ocean and sea ice.

%%%%%%%%%%%%%%%%%%%%

A zonal-mean summary of key climatological variables is shown in Fig.~\ref{fig:zonal_profiles}. The meridional distribution of surface air temperature, precipitation, atmospheric and oceanic heat transport for the Warm state is broadly consistent with contemporary observations. In particular, there is weaker poleward ocean heat transport (OHT) in the Southern Hemisphere than in the Northern Hemisphere, a marked feature of the contemporary climate. The vertical pink lines indicate the latitude of the annual mean sea-ice edge. In today's climate, the southern ice edge reaches $\sim 55$ ° S at its winter maximum and the northern sea-ice edge reaches approximately 45°N at its winter maximum along the eastern coast of Canada. Thus, our Warm solution produces excessive sea-ice cover in the Southern Hemisphere while remaining broadly consistent with observations in the Northern Hemisphere (Fig.~\ref{fig:GLOBAL_WINDS}), as discussed in ~\citet{Moinat_2026}. The combination of precipitation and surface air temperature (SAT) supports substantial Net Primary Productivity (NPP) in the Amazon and Congo basins (Fig.~\ref{fig:FOUR_PLANETS_ORTHO}), consistent with the distribution of tropical forests observed today. The NPP gradually decreases towards the poles, where tundra and boreal forests emerge.

In summary, our Warm solution reproduces the main features of the pre-industrial climate at first order, with a consistent and coherent representation of the major climatic components.

%%%%%%%%%%%%%%%%%%%%

%%%%%%%%%%%%%%%%%%%%%%
\subsection{Hot state}

\begin{figure}[ht!]
\includegraphics[width=\textwidth]{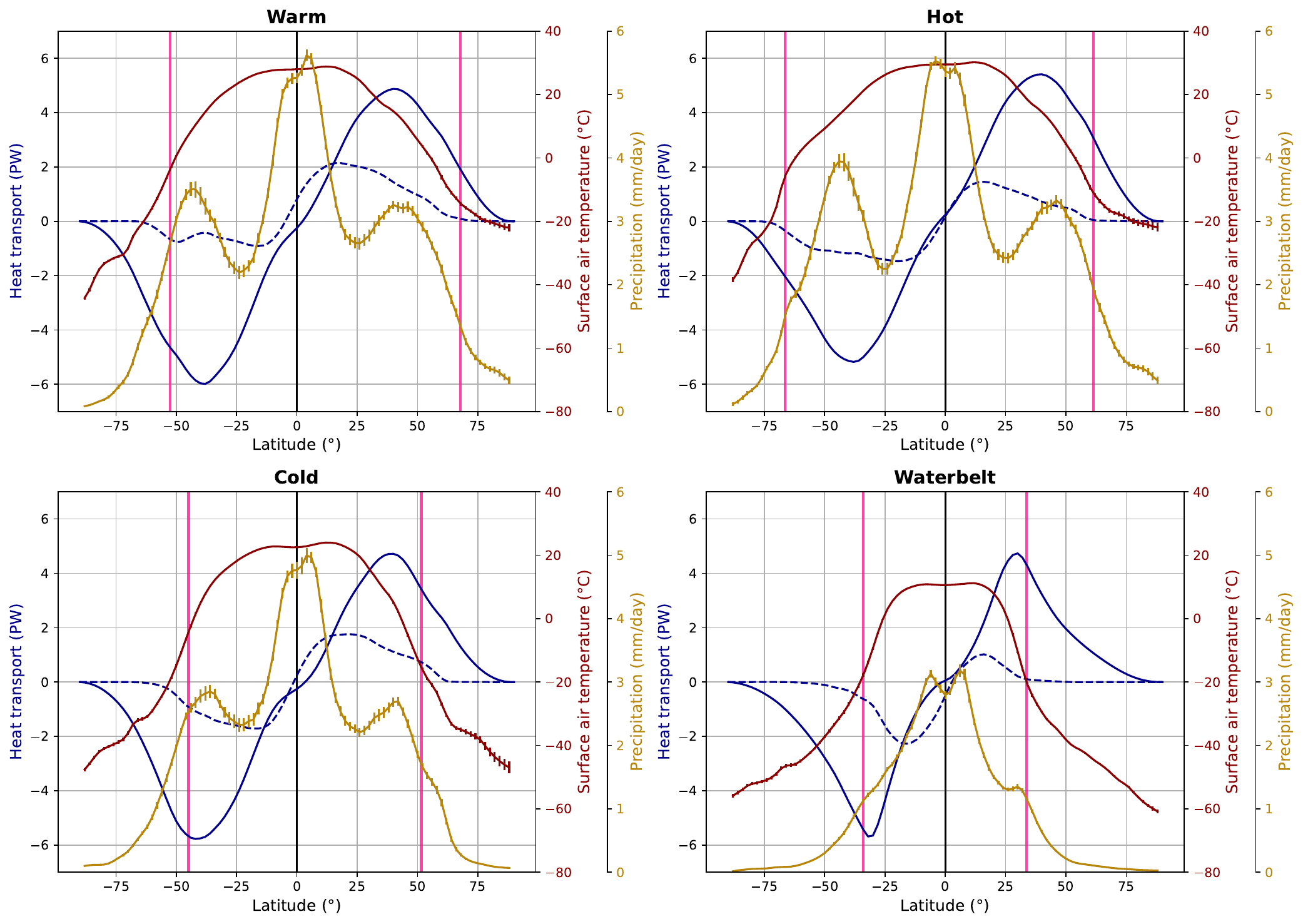}
\caption{Zonal-mean profiles of surface air temperature (red in °C), precipitation (yellow in mm/day) and heat transport (blue in PW with full lines for the atmosphere and dashed lines for the ocean surface) for the Warm (top left), Hot (top right), Cold (bottom left) and Waterbelt (bottom right) states. The pink lines denote the annual mean sea-ice edge for the Northern and Southern Hemispheres. For the Northern Hemisphere, the longitudes used to compute the average are those located in the Pacific and Atlantic basins.}
\label{fig:zonal_profiles}
\end{figure}

In the Hot state, the global-mean SST is 2.86°C warmer than in the Warm state, where this warmth is evident in Figs.~\ref{fig:FOUR_PLANETS_ORTHO}, \ref{fig:Figure3} and \ref{fig:zonal_profiles}. In our Hot solution at a CO$_2$ concentration of 280~ppm, the extent of sea ice in the Southern Hemisphere has reduced markedly, by a factor of approximately 20, and remains confined largely to the Antarctic coast, even during winter. In contrast, Arctic sea ice extent increases by a factor of approximately 1.3 relative to the Warm state, but this increase is far from sufficient to compensate for the loss of sea ice in the Southern Hemisphere.

The ice sheets respond differently: Greenland loses ice in both extent and volume (Table~\ref{tab:PT_global} and Fig.~\ref{fig:FOUR_PLANETS_ORTHO}), while Antarctica exhibits regional gains in surface mass balance as a result of increased polar precipitation. The warmer climate increases precipitation in the Southern Hemisphere (Fig.~\ref{fig:zonal_profiles}), while temperatures remain sufficiently low for snow accumulation to occur. 
However, when considering the total volume of the ice sheet, the net effect is an equilibrium rise in sea-level of $+1$~m relative to the Warm state under the pre-industrial CO$_2$ concentration.

The atmospheric and ocean overturning circulations shown in Fig.~\ref{fig:Figure3} reveal that a primary characteristic of the Hot state is the collapse of the AMOC (see also Fig.~\ref{fig:AMOC-PMOC}). A positive PMOC cell appears in the SH, while the negative PMOC cell strengthens and extends towards the NH (Fig.~\ref{fig:AMOC-PMOC}), 
with oxygenated waters reaching the abyss predominantly from the Antarctic sector (Fig.~\ref{fig:oxygen}). The Southern Ocean is warmer and saltier, while the Arctic is colder and fresher than the Warm state (Figs.~\ref{fig:Figure3} and \ref{fig:sss_humidity}). In addition, the Atlantic–Pacific seesaw (Fig.~\ref{fig:AMOC-PMOC}) is absent, since the Pacific OHT is comparable to that in the Warm state, whereas the Atlantic OHT is close to zero. Despite the collapse of the AMOC, the Bering Strait remains open, thus preventing the formation of deep convection in the North Pacific~\citep{hu2012,saenko2004} (see Figs.~\ref{fig:AMOC-PMOC} and ~\ref{fig:oxygen}).

In contrast to our Warm solution and the present climate, the upper cell of the global overturning circulation of the ocean, which is currently driven primarily by northern sinking, is entirely absent and is replaced by an anticlockwise cell sourced from the south (Fig.~\ref{fig:Figure3}). These changes in oceanic circulation reduce northward OHT in the Northern Hemisphere, while increasing southward transport in the Southern Hemisphere (see Fig.~\ref{fig:zonal_profiles}), consistent with the bipolar seesaw concept of \citet{StockerandJohnses_2003}. Both atmospheric heat transport (AHT) and OHT approach zero directly at the equator (Fig.~\ref{fig:zonal_profiles}), indicating a strong reduction of inter-hemispheric heat transport in the Hot solution~\citep{marshallITCZ2014}. This is reflected in the much greater symmetry of temperature, salinity, and other climatic properties about the equator compared to the Warm solution.

This change in oceanic circulation and SST distribution leads to a southward shift of Hadley cells and, consequently, of the ITCZ (Figs.~\ref{fig:Figure3} 
and \ref{fig:zonal_profiles}). The ITCZ changes associated with AMOC collapse were first observed in \citet{Broccoli_2006}; see also \citet{marshallITCZ2014} and \citet{Jackson2015AMOC}. Moreover, possibly in response to changes in OHT, AHT increases in the Northern Hemisphere and decreases in the Southern hemisphere, consistent with the Bjerknes compensation hypothesis \citep{bjerknes1964}. This is associated with an intensification of the Northern Hemisphere Hadley cell and a weakening of the Southern Hemisphere cell (Fig.~\ref{fig:Figure3}). 
The distribution of surface humidity shifts southward, resulting in a more humid Southern Hemisphere and a drier Northern Hemisphere (Fig.~\ref{fig:sss_humidity}). Surface winds strengthen in the Northern Hemisphere   
(Fig.~\ref{fig:GLOBAL_WINDS}).

These substantial changes in atmospheric circulation lead to a reorganization of vegetation, as reflected in the NPP distribution shown in Fig.~\ref{fig:FOUR_PLANETS_ORTHO}. The main changes are associated with the southward shift of the ITCZ, which leads to a decrease in NPP in northern Amazonia and tropical central Africa, while an increase is found in southern Amazonia (Fig.~\ref{fig:npp_diff_map}). 
The impact of the collapse of the AMOC in northern Amazonia has been documented in paleoclimate records~\citep{Akabane2024AMOC}, while the increase in the gain of the NPP in southern Amazonia is consistent with other simulations~\citep{Ciemer2021}. On a global scale, NPP decreases in Europe, associated with a colder climate, and in Africa, where extremely dry and warm conditions limit vegetation productivity, thus limiting vegetation growth. An overall increase in NPP occurs over Australia, while NPP changes over North America and Asia are comparatively small.  

In summary, the Hot solution represents a climate regime with a collapsed AMOC at the same CO$_2$ concentration as the {\it piControl}, with a mean surface air temperature $\sim 3~^\circ$C higher.

%%%%%%%%%%%%%%%%%%%%%%%
\subsection{Cold state}

The Cold state at a CO$_2$ concentration of 220~ppm resembles a glacial climate with a global-mean SAT of around 5$^\circ$C. This state is mainly characterized by the presence of massive ice sheets over the North American continent and on Eurasia and expanded sea-ice cover at both poles, reaching over the western North Atlantic subtropical gyre and equatorward of the Drake Passage (positions of the sea-ice edges are shown in Figs. \ref{fig:FOUR_PLANETS_ORTHO} and \ref{fig:GLOBAL_WINDS}). As a consequence, sea level is approximately 162~m lower than that of the Warm state, in the range of proxy data for the glacial period~\citep{Clark_2025}. The development of these ice sheets is concomitant with the globally colder climate, as shown in Figs.\ref{fig:FOUR_PLANETS_ORTHO} and \ref{fig:zonal_profiles}. Both poles are similarly cold, and the abyssal ocean is colder than in the Warm state, with deep waters sourced primarily from the Antarctic region (Fig.~\ref{fig:Figure3}).

Atmospheric temperatures are also more symmetrically distributed about the equator. The upper cell of the GMOC is shallower and weaker, while the deep cell is much stronger and of greater vertical extent, pushing far into the abyss of the Northern Hemisphere oceans (Fig.~\ref{fig:Figure3}). 
This behavior is consistent with the results of \citet{Ferrari_2014}. 
In the Southern Hemisphere, the presence of the negative overturning cell is due to the negative overturning circulation in the Pacific (Fig.~\ref{fig:AMOC-PMOC}).
In our simulations, the extent of sea ice in the Southern Hemisphere increases from 22.6 $\cdot 10^6$ km$^2$ in the Warm state to 43.1 $\cdot 10^6$ km$^2$ in the Cold state, consistent with the description in~\citet{Ferrari_2014}. 

The zonal profiles and maps obtained in our simulation are comparable to those produced in the Paleoclimate Modelling Intercomparison Project version 4 (PMIP4) experiments~\citep{PMIP3_4}. In particular, there is a strong asymmetry in the differences between the zonal temperature profiles of the Warm and Cold states, as also observed in all models in PMIP4. This asymmetry can be explained by the massive ice sheet in the Northern Hemisphere, whereas in the south, the ice sheet is constrained by the size of Antarctica. 

 Salinity is generally higher than in the Warm state (by 1.4~psu at the surface) due to a 162 m lower sea level, with prominent subtropical salt tongues, especially in the Atlantic (Fig.~\ref{fig:sss_humidity}). The AMOC intensifies relative to the Warm state, although proxy-based estimates provide no consensus on its strength during colder climates~\citep{lippold2012,Gu2020}. Most of the models in PMIP4 also exhibit an increase in the intensity of AMOC, although its magnitude varies considerably, from +1~Sv to +20~Sv at 30°N. Some models display a deeper positive overturning cell; however, there is no consensus on this behavior~\citep{PMIP3_4}. 
 Possibly as a compensating response~\citep{ferreira2018_AMOC-PMOC}, the negative cell of the PMOC in the Pacific has extended and intensified. Salty abyssal waters are transported from the south towards the north. There is also an indication of an enhanced (relative to Warm) clockwise PMOC in the upper km of the north Pacific (Fig.~\ref{fig:AMOC-PMOC}). Oxygenated water is transported from the north to the south in the Atlantic basin (Fig.~\ref{fig:oxygen}). Extensive sea-ice cover around Antarctica limits air-sea gas exchange so that, unlike Warm, the main source of deep oxygen is from the polar north. There is some evidence of oxygen subduction from the south in the Pacific basin, but with reduced intensity.

The reduction in runoff, associated with the decrease in precipitation shown in Fig. \ref{fig:zonal_profiles}, contributes to an intensification of the positive AMOC overturning cell and the negative PMOC overturning cell. This is reflected in the OHT, where we have an increase in the south and a pattern broadly similar to that of the Warm state in the north, due to the competing effect of AMOC and PMOC~\citep{saenko2004,hu2012} (Figs.~\ref{fig:Figure3} and \ref{fig:AMOC-PMOC}). This increase in AMOC is also associated with changes in the extent of sea ice. In this colder climate, the extent of sea ice has increased globally at both poles, except in the Atlantic basin of the northern hemisphere, where the extent of sea ice in the convection region remains comparable to that of the warm state (Figs.~\ref{fig:FOUR_PLANETS_ORTHO} and \ref{fig:GLOBAL_WINDS}). 

The development of these extensive ice sheets leads to a global intensification of Hadley cells, including both Northern and Southern cells (Fig.~\ref{fig:Figure3}), where the winds are generally stronger in Cold than in Warm (Fig.~\ref{fig:GLOBAL_WINDS}) and the atmosphere is substantially drier (Fig.~\ref{fig:sss_humidity}). The winds are strongly perturbed by the presence of mid-latitude ice-sheets. 
The ITCZ is located closer to the equator than in the Warm state and the maximum peak equatorial precipitation is narrower (Fig.~\ref{fig:zonal_profiles}). 
In addition, sea-ice expansion along the eastern coast of the North American continent is driven by northerly winds over the ice sheet, which transport cold air into the region and stabilize the edge of the sea ice (Fig.~\ref{fig:GLOBAL_WINDS})~\citep{BroccoliManabe1987}.
Moreover, the positive polar cell in the Northern Hemisphere becomes stronger than in the Warm state, owing to the growth of the Northern Hemisphere ice sheet, and the jet stream intensifies in the Northern Hemisphere (Fig.~\ref{fig:zonal_wind_verticsal_4states}). Lower temperatures and the associated decrease in specific humidity (Fig.~\ref{fig:sss_humidity}) lead to a global decrease in precipitation, particularly in the tropics and mid-latitude regions (Fig.~\ref{fig:zonal_profiles}), in agreement with the results in~\citet{PMIP3_4}. 
All these changes in the atmosphere cause an overall decrease in the NPP (Fig.~\ref{fig:FOUR_PLANETS_ORTHO} and ~\ref{fig:npp_diff_map}), with equatorward migration of tundra and boreal forest (Fig.~\ref{fig:biome_map}). 

In summary, our cold state captures key features of typical glacial climates. There is a global cooling of $\sim 9$ °C with respect to the Warm state, and a tropical cooling of $\sim 3$ °C falls within the consensus ranges \citep{Clark_2025}. The expanded ice sheets and sea ice, stronger and drier winds, intensified Northern Hemisphere Hadley Cell, equatorward-displayed ITCZ with a narrower precipitation peak maximum, higher ocean salinity associated with lower sea level, and expanded southern-sourced deep-water formation are all broadly consistent with proxy evidence and previous model simulations. 

%%%%%%%%%%%%%%%%%%%%%%%%%%%
\subsection{Waterbelt state}

The Waterbelt state is characterized by sea-ice cover that extends to approximately 30$^\circ$ latitude in both hemispheres, and by ice sheets in the northern hemisphere that extend equatorward to approximately 30 $^\circ$N, with a mean global SAT of $-14^\circ$C (Table~\ref{tab:PT_global} and Fig.~\ref{fig:FOUR_PLANETS_ORTHO}). 
The existence of this state is primarily associated with the convergence of ocean heat transport near the edge of the sea ice, as shown in Fig.~\ref{fig:zonal_profiles}. The heat transport convergence supplies sufficient heat to the ice edge to counteract the ice-albedo feedback and prevent further equatorward expansion of sea ice~\citep{Rose2015}.

The substantial reduction in sea level results in a considerably saltier ocean, both throughout the depth profile and at the surface, with SSS increasing by 2.35 psu relative to the Warm state.

The GMOC is characterized by a strong negative overturning cell, while wind-driven cells dominate near the sea-ice edge. In addition, OHT is more strongly negative in the Southern Hemisphere than positive in the Northern Hemisphere. This asymmetry is mainly associated with the strong negative overturning cell due to the AABW, together with the disappearance of the NADW, a behavior similar to that reported by \citet{Obase_2026}.

Both OHT and AHT are substantially reduced relative to the Warm state, due to weaker atmospheric overturning cells  (Fig.~\ref{fig:Figure3}). This reduction in heat transport is associated with lower precipitation, reduced specific humidity, and the corresponding reduction in NPP. In addition, the Waterbelt state is characterized by a strong meridional SAT gradient in both hemispheres (Table~\ref{tab:PT_global}). In contrast to the Hot and Cold states, substantial changes occur in ocean currents and surface wind patterns, with pronounced intensification in the vicinity of the extensive ice sheets (Fig.~\ref{fig:GLOBAL_WINDS}). 

\section{Role of wind, eddies and ocean heat transport in setting the ice edge}
A central feature of climates that support multiple equilibria is the location of the ice edge in each state. Because ice-albedo feedback is extremely strong, even modest shifts in that edge can lock the system into one attractor or another and thereby control the stability of the entire configuration. Sea ice and ocean circulation respond so sensitively to the prevailing winds that the connection to wind forcing is likely to be key in setting where the ice edge lies.

\citet{enomoto1990} linked the edge of the winter Antarctic sea ice with the
atmospheric convergence line around Antarctica, the latitude at which the mean zonal wind of the surface changes from polar east to mid-latitude westerlies. They argued that southward of the zero zonal-wind line, the associated ocean Ekman transport is southward and {\it vice versa}. This favors southward transport or convergence/compaction of ice near the continent. Thus, if nothing else is at play, one might expect the winter ice edge to align with the zero wind line because that line marks the transition in the large-scale wind field that controls the direction and magnitude of ice advection.

The first line of Fig.~\ref{zonal_stress_SH} plots the annual-mean sea-ice edge found in our four steady states along with the zonal component of the wind stress. We observe no general correspondence between the ice edge and the zero wind line. However, note that the  argument in \citet{enomoto1990} is kinematic: the wind reversal changes the direction of ice drift and, therefore, the net northward or southward displacement of the pack. However, our sea-ice model does not include ice dynamics and is purely thermodynamic --- it cannot directly capture the mechanism. That said, the observational study of \citet{goosse2025} shows that the ice edge covaries much more closely with the polar front of the Antarctic Circumpolar Current (ACC) than with the zero wind line. The mechanism they invoke is thermodynamic. The Polar Front (and the ACC path more generally) influences the latitude at which warmer water and the associated southward heat fluxes keep the mixed layer from reaching the freezing point.

The discussion of \citet{goosse2025} is consistent with that of \citet{Rose_Marshall_2009}, who argue that the edge of the sea ice lies at latitudes where the ocean brings in heat, and the OHT becomes small and eventually vanishes at the edge of the ice. The ice itself acts as an insulator, blocking air-sea heat exchange. We now attempt to bring together these various strands into a diagnostic approach that can be tested in our various equilibrium solutions.

\subsection{The sea-ice edge and surface residual flow}
In the presence of mesoscale eddies, which are ubiquitous in the southern
ocean and parameterised in our model, the ocean advects tracers, including temperature and heat with the residual velocity~\citep{marshallRadko2003}. At the surface of the ocean the meridional component of the residual current can be written:
\[
v_{\text{res}}\ =v_{\text{Ek}}\ +v^{\ast } 
\]
where $v_{\text{Ek}}$ is the Ekman transport and $v^{\ast }$ is the
eddy-induced velocity (bolus) that depends on the stratification of the ocean and therefore the frontal position. 
We argue that in the Southern Ocean it is the line where $v_{\text{res}}$ vanishes that matters, rather than the Eulerian mean $v_{\text{Ek}}$. 

The latitude at which $v_{\text{res}}$ vanishes (or changes sign) is a
plausible dynamical boundary setting the ice edge because the residual circulation is the key to meridional heat transport. The location where residual meridional flow vanishes will likely coincide with transitions in surface heat flux and the ability of the ocean to melt
ice from below.

In Supplementary Sect.~S3, we summarize elements of residual-mean theory~\citep{marshallRadko2003,FerreiraMarshall2006} which provide the dynamical framework which we use to diagnose the surface meridional residual flow. Two key stresses are at play: the applied zonal wind stress $\tau _{x}^{w}$ due to the prevailing westerly winds and a (quasi-balancing) mesoscale eddy stress $\tau _{x}^{e}$, which is parameterised in our model. This carries the wind stress down the water column, where it is balanced by topographic form drag at the bottom. The net stress on the upper layer of the model, floating on which is the sea ice, is then given by $\tau _{x}^{w}-\tau _{x}^{e}$. This is largely balanced by Coriolis accelerations, and so the direction of the meridional flow at the top of the ocean is set by the sign of $\tau _{x}^{w}-\tau _{x}^{e}$.

\begin{figure}[ht!]
\centering
\includegraphics[width=\textwidth]{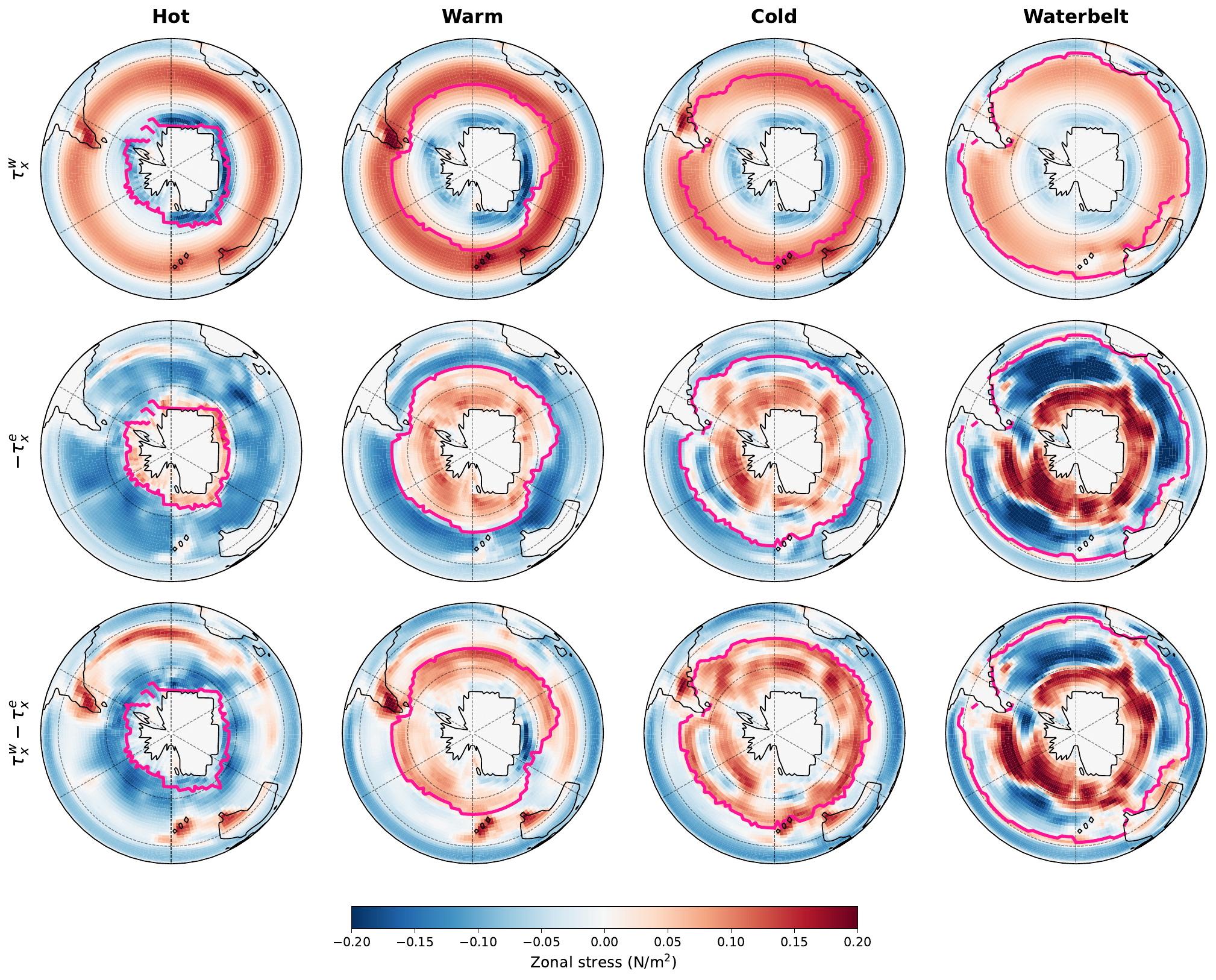}
\caption{The mean annual zonal component of wind stress, $\tau_x^w$, of minus the eddy stress, $-\tau_x^e$, and of their sum, $\tau_x^w-\tau_x^e$, in the Southern Hemisphere, expressed in $\mathrm{N}\,\mathrm{m}^{-2}$. The pink line represents the annual mean sea-ice edge.}
\label{zonal_stress_SH}
\end{figure}

Fig.~\ref{zonal_stress_SH} (first row) shows the zonal component of the
mean annual wind stress, $\tau _{x}^{w}$, with the annual-mean position of the edge of the ice marked with the pink line. The familiar pattern of easterly trade winds, middle latitude westerlies and polar easterlies is evident in each of the climates attesting to the robustness of general global wind pattern across very different climates. There is no clear association of the ice edge with zero-crossings of the wind, except in the Waterbelt where the ice pushes
down to the polar edges of the trade wind belt where easterlies switch over
to westerlies.

Fig.~\ref{zonal_stress_SH} (second row) shows the (minus) zonal component of the
eddy stress, $-\tau _{x}^{e}$, evaluated at a depth of $20~\unit{m}$, at the
bottom of the upper layer of the model. As can be seen in Eq.~(\ref%
{res_mtm_x_compt}), $-\tau _{x}^{e}$ can be thought of as an
eastward-directed  force. In general, eddy stresses are weak in the tropics, but typically $-\tau _{x}^{e}$ has the opposite sign to $\tau _{x}^{w}$ at
higher latitudes: eddies tend to oppose the action of the wind. 
Note
that in the Southern Ocean, equatorward of the ACC, the wind (eddies) drives
meridional residual flow equatorward (polewards), and {\it vice versa} poleward of
the ACC. The eddy stress has a zero crossing at the ice edge because, as we
shall see, the doming of fresh water up towards the ice edge reverses the
sign of the near-surface meridional buoyancy gradient and hence the sign of the eddy velocity $
\mathbf{v}^{\ast }$.

Fig.~\ref{zonal_stress_SH} (third row) shows the net zonal stress acting on the
surface ocean, $\tau _{x}^{w}-\tau _{x}^{e}$, the sum of the wind and (minus) eddy components. In Warm and Cold, we
observe a net zonal eastward stress to the south of the ice edge, driving
surface residual flow northward and a net (generally) westward stress north
of the ice edge, driving residual flow southward. The edge of the ice lies in a
 stagnation region, where $v_{\mathrm{res}}\simeq 0$, consistent with our
hypothesis. 
In Hot, there is a net westward
stress acting on the ocean all around Antarctica that acts to drive water
and ice toward the coast: indeed, the ice edge is very close to the Antarctic
coast. In the Waterbelt, the ice edge lies on the polar flanks of
the tropics close to the zero wind and a sign change in the net forcing.
It is interesting that even in the northern hemisphere, away
from the circumpolar regime of the southern ocean, the Pacific ice edge also
resides close to stagnation latitudes (Fig.~\ref{zonal_stress}). 

\begin{figure}[ht!]
\centering
\includegraphics[width=\textwidth]{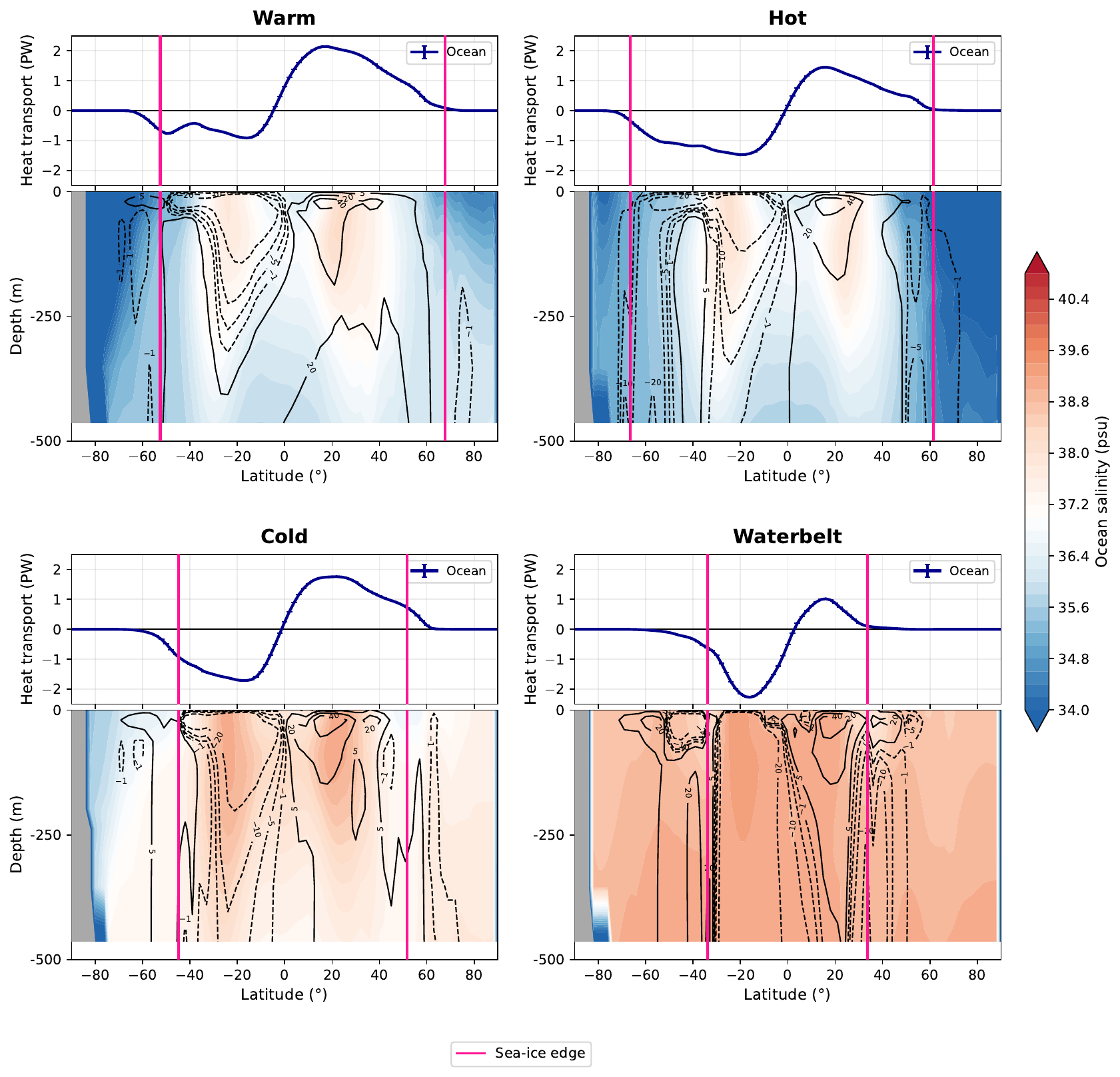}
\caption{Ocean heat transport (PW) is shown in the upper panel for each state. The lower panel shows ocean salinity (psu) using a red–blue colour scale, with the residual streamfunction represented by contours. Solid contours indicate clockwise circulation, while dashed contours indicate counter-clockwise circulation, for each of the four equilibrium states. The pink vertical line denotes the annual-mean latitude of the sea-ice edge.}
\label{residual_streamfunction}
\end{figure}

Fig.~\ref{residual_streamfunction} shows the residual overturning circulation,
zoomed into the top 500$~\unit{m}$ of the ocean.  
The sense of overturing is
consistent with the surface forcing maps shown in Figs.~\ref{zonal_stress_SH} and \ref{zonal_stress}. Note the shallow overturning cells in the upper ocean and the close
association of the ice edge with a change in the sign of the residual
overturning cells where $v_{\mathrm{res}}\simeq 0$.
Fig.~\ref{residual_streamfunction} shows how the presence of ice
freshens the surface ocean creating a halocline. The salinity field domes up
toward the surface at the ice edge and plays a role in reversing the sign of
the residual flow across the ice edge.

The residual circulation is intimately associated with meridional OHT, as shown in the upper part of the panels in Fig.~\ref{residual_streamfunction}, its divergences and thence air-sea fluxes. Fig.~\ref
{fig:Figure9} shows the net air-sea flux. There is generally a
convergence of OHT toward the ice edge, suggesting that heat transport by the
ocean plays an important role in arresting equatorial encroachment of the
ice. Note also that the air-sea heat flux vanishes in regions of ice cover
because sea ice is insulating, at least in our model.

%%%%%%%%%%%%%%%%%%%%%%

%%%%%%%%%%%%%%%%%%%%%
\subsection{Connection to extensions of the Budyko--Sellers system}
Classical EBMs have a smooth, atmosphere-like heat transport (a single
hemispheric bump). Ice-albedo feedback then leaves only two robust
attractors: a small polar cap and a snowball. A mid-latitude ice edge is
unstable (the small-ice-cap instability). \citet{Rose_Marshall_2009} put two ocean ingredients into the EBM: ice insulates the atmosphere from the ocean, and OHT has a meridional structure set by the winds --- gyre/Ekman cells with
heat transport minima (or equivalently, heat transport convergence just
equatorward of those minima). The ice edge can rest at those structured
latitudes. That creates extra stable states (notably a large mid-latitude
cap) and hysteresis: the edge can jump from mid- to high latitudes under a
small change in forcing. The \citet{Ferreira2011} and \citet{Brunetti_2019} MITgcm aquaplanet/ridge integrations realized similar states in
a GCM: Hot (ice-free), Cold, Waterbelt, and Snowball. In
those runs OHT convergence just equatorward of the ice edge is what holds
the Cold/Waterbelt edges.

Our diagnostics can be seen in the same framework: $v_{\mathrm{res}}=0$ is
the kinematic statement of the same structure. Residual meridional flow is
the agent of OHT in the upper ocean:
\[
OHT\sim \rho\, c_{p}\int v_{\mathrm{res}}\,T\,dz.
\]%
A zero of surface $v_{\mathrm{res}}$ is a stagnation latitude for residual
mass and, typically,  
an inflection point or peak in the residual heat-flux
convergence. Heat is carried toward that latitude from the open-ocean side
and is not carried efficiently under the ice (ice also cuts surface stress
and shuts down the local residual cell). That is exactly the
\textquotedblleft preferred latitude/OHT minimum\textquotedblright\ \citet{Rose_Marshall_2009} put in the EBM.

Our two observations -- the edge of the ice at $v_{\mathrm{res}}=0$, and the
 transport of ocean heat toward the edge -- are the GCM-level counterparts of
\textquotedblleft the edge of the ice sits where the ocean converges heat\textquotedblright\, thus compensating the cooling of the ice-albedo and arresting the runaway. Multiple equilibria exist because OHT is not a smooth sinusoid. Structured ocean transport creates discrete latitudes at which the local energy budget at the ice line can close. Ice insulation also matters: once ice is present, the ocean cannot warm the atmosphere through it, so the heat that stabilizes the edge must be delivered at the edge, not under the pack.

In our more complete model, the structure is not an imposed wind-curl
pattern in an EBM; it is the residual circulation that includes $\mathbf{v}%
^{\ast }$ and makes a crucial contribution. That is why the edge tracks 
$v_{\mathrm{res}}=0$ rather than $v_{\text{Ek}}\ =0$. Eddies and the
ice-induced salinity front move the OHT convergence away from the wind-only
line. Heat transport toward the edge is the thermodynamic content of that
residual cell, not just a prescribed OHT profile. In this way, the OHT structure
 sets both preferred latitudes and responds to the ice.

In summary, our more complex model is doing what an ocean-aware
Budyko--Sellers system should do: discrete, heat transport-supported ice
edges instead of a unique polar cap. Identifying those edges with $v_{%
\mathrm{res}}=0$ and with residual heat transport onto the edge is an
appropriate dynamical reading of their OHT-minimum mechanism, and it is
consistent with the residual-mean ice edge argument.

%%%%%%%%%%
\begin{figure}[ht!]
\includegraphics[width=\textwidth]{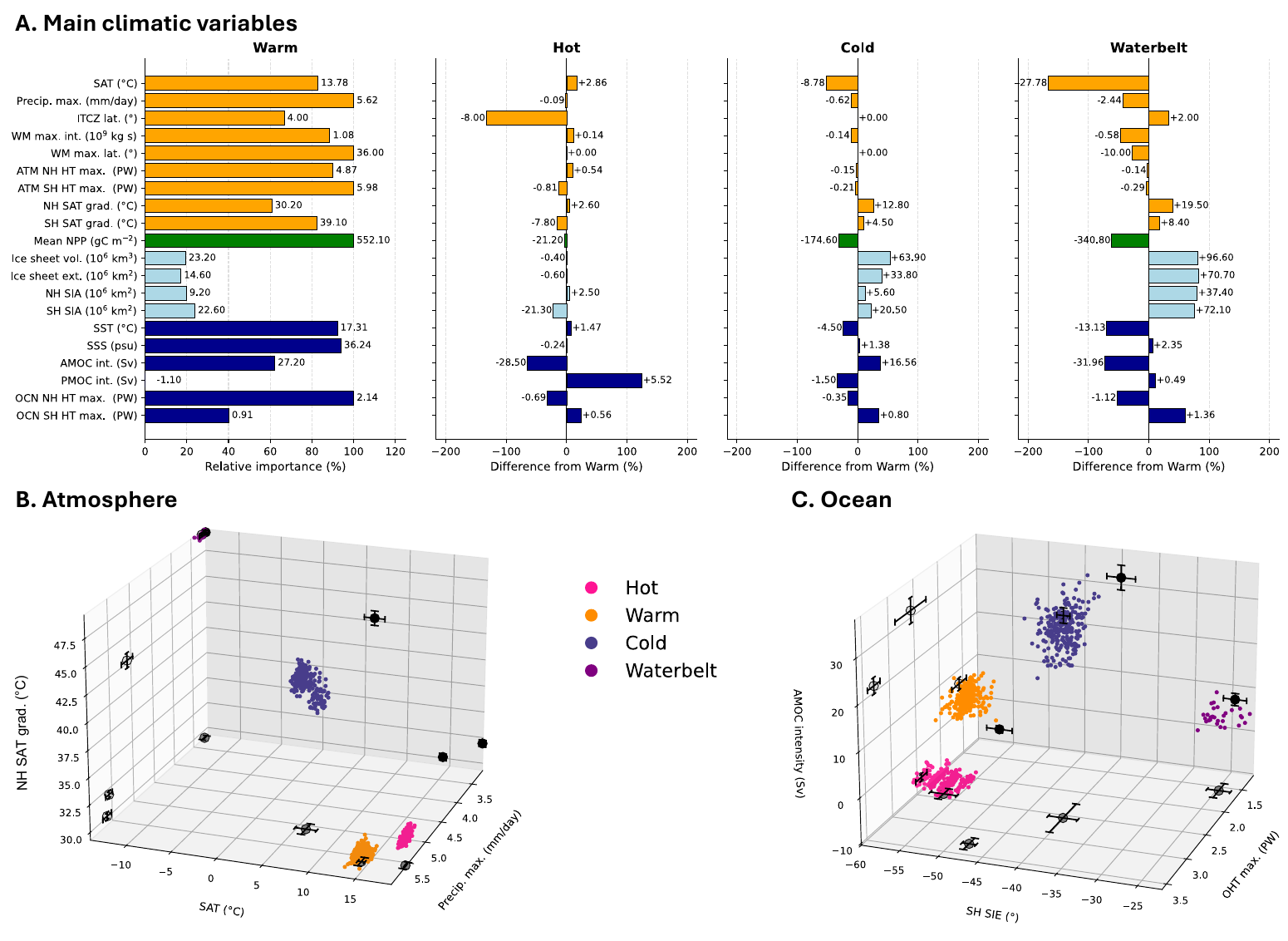}
\caption{A. Summary of the main climatic variables
in the Warm state (first panel), and differences of the other three attractors  with respect to the Warm state. The bar lengths are proportional to the values in the Warm state. The climate variables are classified as: atmosphere (orange), vegetation (green), cryosphere (light blue) and ocean (blue). B. Trajectory of three atmospheric climate variables on the Warm (orange), Hot (pink), Cold (dark blue) and Waterbelt (purple) attractors. C. Trajectory of three oceanic climate variables on the same attractors. The acronyms are defined in the caption of supplementary Table \ref{tab:PT_global}.
}
\label{fig:phasespace}
\end{figure}

%%%%%%%%%%%%%%%%%%%%%%
\section{Discussion and Conclusions}  %% \conclusions[modified heading if necessary]

The five equilibrium solutions (Warm, Cold, Hot, Waterbelt and Snowball) obtained under very similar CO$_2$ forcing demonstrate that the coupled climate model possesses multiple stable states. Their differences arise primarily from strong nonlinear feedbacks involving the Atlantic Meridional Overturning Circulation (AMOC), continental ice sheets, sea ice, and the associated inter-hemispheric and inter-basin energy transport, rather than from differences in greenhouse-gas concentrations, as summarized in Fig.~\ref{fig:phasespace}A. The attractors occupy distinct regions in phase space and represent distinct clusters in the 3-dimensional plots in terms of key atmospheric and oceanic parameters shown in Fig.~\ref{fig:phasespace}B, C.
The resulting climate regimes depend on the organization and balance of the interacting mechanisms between these climatic components constrained by the residual circulation of the ocean, which sets the edge of the sea ice, as we have identified in our simulations. 

The Warm state corresponds to the familiar “AMOC-on” modern climate. The Cold state is an ice-sheet-dominated glacial climate with intermediate global overturning strength. It is characterized by extensive Northern Hemisphere ice sheets and expanded sea-ice cover. The effects of the high albedo and topographic barrier of the ice sheets (the mechanism isolated by \citet{BroccoliManabe1987}) strongly cool the Northern Hemisphere and strengthen the atmospheric circulation. 
Higher mean ocean salinity (from a 162-m lower sea level) and altered freshwater fluxes intensify the AMOC, while the deep, southern-sourced cell expands into the northern abyss. 
The Cold equilibrium is thus maintained by mutually reinforcing greater ice-sheet and sea-ice extent relative to the Warm state, which strengthens the ice-albedo feedback. Feedbacks associated with heat transport and overturning circulation prevent that feedback from becoming degenerate, thus stabilizing a glacial-like climate.

The Hot state lies on the opposite branch of AMOC hysteresis. The northward AMOC cell has collapsed and has been replaced by a southern-sourced overturning circulation. Ocean heat transport therefore reverses its interhemispheric asymmetry, producing the bipolar seesaw described by \citet{StockerandJohnses_2003}: heat accumulates in the Southern Hemisphere while the Northern Hemisphere cools relative to the Warm state. 
Stronger Southern Hemisphere heat transport relative to the Warm state causes sea ice to collapse around Antarctica, removing an important albedo effect and altering the surface heat budget, while sea ice expands modestly in the Arctic, particularly in the Atlantic sector. 
The Hot equilibrium is therefore stabilized by the “AMOC-off” circulation, the associated bipolar temperature seesaw, and the consequent rearrangement of sea ice and ice-sheet mass balance.

In the above three climate regimes, the edge of the sea ice  corresponds to the latitude where the residual velocity becomes null, showing the crucial role of the surface stress associated with both wind and eddies in shaping the ocean heat transport convergence and in limiting the extent of the ice.  

The waterbelt state has sea ice and Northern Hemisphere ice sheets extending to $\sim$30$^\circ$ latitude and a global-mean SAT of more than $27^\circ$C colder than today. The large ice sheets and sea-ice cover are maintained by the convergence of the ocean heat-transport by the subtropical ocean cells out to the latitude where the easterly trade winds give way to middle-latitude westerlies. 
In this climate regime, eddy stress does not play a role, and the sea-ice edge is thus governed by the zero of the Ekman velocity component. 

The existence of five different climate regimes under the same pre-industrial radiative forcing in the present-day continental configuration needs to be confirmed by other modeling groups through multimillennial coupled simulations, albeit this may be too computational expensive for some model setups. Based on the present analysis, the minimal ingredients required for multistability are a fully dynamical ocean that includes eddy parameterization, as well as sea ice and ice sheets that are consistent with the underlying climate regime. 

Understanding the dynamical structure of Earth’s climate system with models that incorporate millennial-timescale processes is crucial to identify the mechanisms that stabilize these states and locate them in phase space, particularly because transitions between them can occur.
Transitions between alternative climate regimes would require finite-amplitude perturbations, such as changes in CO$_2$ or astronomical forcing, capable of moving the system from one basin of attraction to another.
The Warm and Cold states may be relevant to glacial–interglacial transitions over the last million years, which are believed to be driven by the combined effects of Milankovitch cycles and changes in atmospheric CO$_2$. The Hot state may be part of a hothouse branch that could be reached under continued anthropogenic forcing. 
More work is required to determine the range of existence of each climate regime and to assess their implications for the past and possible future trajectories of the Earth’s climate.

%%%%%%%%%%%%%%%%%%%%%%%%%%%%%%%
\section{Modelling Framework and Strategy}

\subsection{\it BIG-MITgcm}
\label{sec:model}
We now briefly describe {\it BIG-MITgcm}, the coupled framework used to perform our simulations. A more detailed description and comparison to observations can be found in~\citet{Moinat_2026}. The dynamical core is provided by the MIT General Circulation Model (MITgcm)~\citep{marshall_finite-volume_1997,Marshall1997b,Marshall2004,adcroft_implementation_2004}. We use a coupled MITgcm setup that includes atmosphere, ocean, land and thermodynamic sea ice on the same cubed-sphere grid with 32 $\times$ 32 points per face (denoted CS32), which is equivalent to a horizontal resolution of 2.8°, or roughly 250 km. This enables us to explicitly capture synoptic variability in the atmosphere, but ocean weather systems must be parametrized. We employ a Gent\&McWilliams-type parameterization of ocean eddies~\citep{Gent1990} and a KPP model of ocean mixing~\citep{Large2004}. The ocean has 25 non-uniform levels ranging from 20 m at the surface to 1300 m at depth. The atmospheric physical parameterizations are implemented by the SPEEDY module~\citep{Molteni2003}, with 5 vertical layers, one in the atmospheric boundary layer, one in the stratosphere and three in the troposphere. The thermodynamic sea ice package THSICE is based on~\citet{Winton2000}: there is no representation of sea ice dynamics.  The simulations are run for the present-day orbital configuration and the solar constant (1365.4 W \, m$^{-2}$). In addition, we used the GCHEM, DIC and PTRACERS packages \citep{Follows_2006_biogeochem} to represent dissolved inorganic carbon (DIC), dissolved organic phosphorus (DOP), alkalinity, phosphate, and oxygen tracers for each state. To minimize the computational cost of including these processes, the method described in \citet{RagonPT} is used.
%\citet{Berger}. 

The equilibrium vegetation distribution is obtained using the BIOME4 model~\citep{Kaplan2003}. It is driven by the monthly-averaged surface air temperature, solar insolation, precipitation, and the minimum monthly temperature provided by MITgcm. BIOME4 reconstructs the most probable biome at each spatial point by simulating a competition among twelve functional types of plants. There are a total of 27 different biomes that range from the tundra to tropical forests to the desert. The albedo and vegetation fraction, estimated as in~\citet{Plioprojet} for each biome, are then returned to the coupled MITgcm for further iteration.

The global ice sheet model is MITgcmIS~\citep{Moinat_2026} and is used to reconstruct the extent and thickness of continental
ice not necessarily confined to high latitudes. MITgcmIS makes the shallow-ice approximation to model ice movement,  employs the positive degree day method~\citep{TSAI_RUAN_2018} to compute the surface mass balance, and includes a lapse rate correction. This description of ice-sheet dynamics neglects km-scale processes and allows us to capture orographic effects and the ice-albedo feedback, which are the dominant processes at a resolution of 2.8°. MITgcmIS is directly implemented on the same CS32 grid of the dynamical core. It is compared with observations in~\citet{Moinat_2026}.

The last component of the coupled setup is the runoff model {\it pysheds}~\citep{bartos_2020}. At each location on the land, a surface-water flow is estimated and a runoff routing map is computed. This is updated at each cycle of the coupled system following the changes in topography associated with ice-sheet growth and retreat.

The asynchronous runs of MITgcm (which includes atmosphere, ocean and sea ice), BIOME4, MITgcmIS and {\it pysheds} constitute the coupled framework which we call {\it BIG-MITgcm}. The system is asynchronously cycled until convergence is achieved between all climate components, as detailed in~\citet{Moinat_2026}.

%%%%%%%%%%%%%
\subsection{Mapping out climate states}

To map out the steady states of the model climate, several methods can be used, as described in~\citet{Brunetti_2019, Ragon_2023}. The reference simulation is the pre-industrial climate described in~\citet{Moinat_2026} corresponding to an atmospheric CO$_2$ concentration of 280~ppm, which is our \textit{piControl}. This steady state is called Warm and, as we shall see, corresponds to an  `interglacial' state.

In the present study, two methods are used to explore phase space and identify the attractors. The first consists of gradually increasing or decreasing CO$_2$ (in increments of $0.1$~ppm/yr), as shown schematically in Fig.~\ref{fig:state_tracking_method}. This maps out the stable branch associated with, for example, the Warm attractor under increasing and decreasing CO$_2$ and identifies the location of the so-called global tipping points, where the attractor loses stability and the system transitions to another attractor. An example of this approach for a coupled aquaplanet configuration is provided in \citet{moinat}. Once a new attractor is identified, the coupled system is integrated to convergence, described in Sect.~\ref{sec:model}, with ice sheets and vegetation cover consistent with the new climate conditions. The branch of the new attractor can then be explored using the same technique. The second method is to construct different initial conditions by changing, for example, the ocean temperature distribution, as described in \citet{Brunetti_2019}. This allows one to explore different regions of phase space seeking alternative attractors. In both methods, at each point in phase space the coupled model is integrated out until convergence is achieved among all climate components over multi-millennial timescales. The attractors presented here are performed under the same radiative forcing (solar constant of 1365.4~W \, m$^{−2}$ and CO$_2$ content in the range of 220-280~ppm). 

Mapping out relevant regions of phase space to convergence is computationally expensive. In the present study, approximately 20,000 simulated years were required for the four additional attractors (the warm {\it piControl} was already obtained in \citet{Moinat_2026}): roughly four simulations of 5000 yr each, demanding 1000 simulated years per week on a dedicated cluster. This does not include the additional computational cost associated with the asynchronous coupling of the ice-sheet and vegetation components, which is relatively inexpensive, and with the exploration of the stable branches shown in Fig.~\ref{fig:state_tracking_method}. 

%% The following commands are for the statements about the availability of data sets and/or software code corresponding to the manuscript.
%% It is strongly recommended to make use of these sections in case data sets and/or software code have been part of your research the article is based on.

\dataavailability{The current version of BIG-MITgcm is available from the project website https://doi.org/10.5281/zenodo.18723952 \citep{Moinat_dataset2025} under the license Creative Commons Attribution 4.0 International. The data to reproduce the plots for all the simulations presented in this paper are available at https://doi.org/10.5281/zenodo.23981966, \citep{moinat_2026_zenodo}} %% use this section when having only data sets available

\authorcontribution{MB acquired funding and designed the research framework. LM ran the simulations with the help of MB, and made the plots. JM developed the  methodology based on the residual-mean theory. All the authors analysed the simulation results, wrote the manuscript and reviewed the paper.} %% this section is mandatory

\competinginterests{The authors declare that they have no competing interests.} %% this section is mandatory even if you declare that no competing interests are present

%\disclaimer{TEXT} %% optional section

\begin{acknowledgements}
The authors thank Jean-Michel Campin of MIT for useful discussions on the simulation setup. LM and MB acknowledge the financial support from the Swiss National Science Foundation (Sinergia Project no. CRSII5-213539). JM acknowledges endowed funds from MIT that helped the collaboration. The simulations were performed on high-performance clusters at the
University of Geneva. \end{acknowledgements}

%\codedataavailability{TEXT} %% use this section when having data sets and software code available

%\sampleavailability{TEXT} %% use this section when having geoscientific samples available

%\videosupplement{TEXT} %% use this section when having video supplements available

%\noappendix       %% use this to mark the end of the appendix section. Otherwise the figures might be numbered incorrectly (e.g. 10 instead of 1).

%% Regarding figures and tables in appendices, the following two options are possible depending on your general handling of figures and tables in the manuscript environment:

%% Option 1: If you sorted all figures and tables into the sections of the text, please also sort the appendix figures and appendix tables into the respective appendix sections.
%% They will be correctly named automatically.

%% Option 2: If you put all figures after the reference list, please insert appendix tables and figures after the normal tables and figures.
%% To rename them correctly to A1, A2, etc., please add the following commands in front of them:

%\appendixfigures  %% needs to be added in front of appendix figures

%\appendixtables   %% needs to be added in front of appendix tables

%% Please add \clearpage between each table and/or figure. Further guidelines on figures and tables can be found below.

%% REFERENCES

%% The reference list is compiled as follows:
\bibliographystyle{copernicus}
\bibliography{example.bib}

\appendix
\newpage
\section{Snowball state}
\label{sec:snowball}

The Snowball state, which is not discussed in
detail in the main text, is included here for completeness. It is obtained for the
same atmospheric CO$_2$ content of 280~ppm as the Hot, Warm and waterbelt states. The Snowball state is characterized
by a global mean surface temperature of approximately $-41^\circ$C
and by a surface that is fully covered by sea ice and continental
ice sheets. Its stability is primarily maintained by the ice-albedo
feedback \citep{Snowball_Pierrehumbert}. The total ice-sheet volume
reaches $169 \times 10^6$ km$^3$, corresponding to a sea-level
reduction of approximately 391~m relative to the Warm state.

\begin{figure}[H]
\centering
\includegraphics[width=\textwidth]{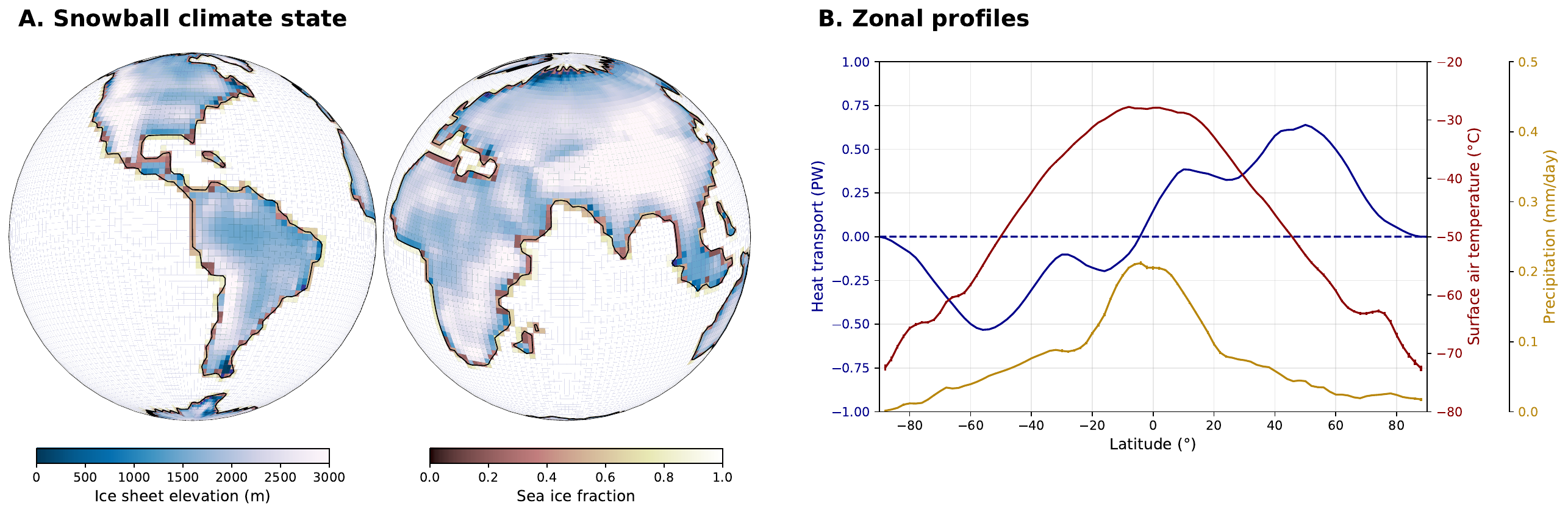}
\caption{A. Representation of the main surface features of the
Snowball state. Ice-sheet elevation is represented by the blue--white
shading over land, while the brown--white shading represents sea ice
over the ocean. SST and NPP distributions are not shown because the
Snowball state is entirely covered by ice over both land and
ocean. B. Zonal profiles of ocean heat transport (blue dashed line),
atmospheric heat transport (solid blue line), surface air temperature
(solid dark-red line), and precipitation (solid dark-yellow line).
Compared with Fig.~6 in the main text, the axes have been adjusted to
highlight the main features of the Snowball state.}
\label{fig:Snowball_combined_figure}
\end{figure}

% ============================================================
% DYNAMICAL FRAMEWORK
% ============================================================

\section{Diagnostic framework}

Our equilibrium solutions have distinctly different ice edge
positions. Our goal is to diagnose the model to see whether there is
a coincidence between the latitude where $v_{\mathrm{res}}=0$ and the
ice edge. A diagnostic framework is provided by residual-mean theory,
which we now outline.

The momentum equation in the planetary geostrophic limit, in which
advection of momentum is neglected, can be written:

\begin{equation}
f\hat{\mathbf{z}}\times\overline{\mathbf{v}}
+\frac{1}{\rho_{\mathrm{ref}}}\overline{\nabla}p
=
\frac{1}{\rho_{\mathrm{ref}}}
\frac{\partial\boldsymbol{\tau}^{w}}{\partial z},
\label{mtm_eq}
\end{equation}
where $f$ is the Coriolis parameter, $p$ is the pressure,
$\rho_{\mathrm{ref}}$ is a constant reference density and
$\boldsymbol{\tau}^{w}$ is the surface wind stress.

The residual velocity is defined as
\[
\mathbf{v}_{\mathrm{res}}
=
\overline{\mathbf{v}}+\mathbf{v}^{\ast}
=
\overline{\mathbf{v}}
-
\nabla\times\boldsymbol{\Psi}^{\star},
\]
where
\begin{equation}
\boldsymbol{\Psi}^{\star}
=
\left(
\frac{\overline{v'b'}}
{\overline{b}_{z}},
-\frac{\overline{u'b'}}
{\overline{b}_{z}},
0
\right)
\label{psi_star}
\end{equation}
is the streamfunction of the eddy-induced circulation~\citep{marshallRadko2003}, with $(u',v')$ the zonal and meridional eddy velocity components, $\overline b$ and $b'$  the mean and eddy buoyancy, respectively. Thus, eq.~(\ref{mtm_eq}) can
be written as
\begin{equation}
f\hat{\mathbf{z}}\times\mathbf{v}_{\mathrm{res}}
+
\frac{1}{\rho_{\mathrm{ref}}}\overline{\nabla}p
=
\frac{1}{\rho_{\mathrm{ref}}}
\frac{\partial
(\boldsymbol{\tau}^{w}+\boldsymbol{\tau}^{e})}
{\partial z}.
\label{res_mtm}
\end{equation}
where
\begin{equation}
\boldsymbol{\tau}^{e}
=
(\tau_x^e,\tau_y^e)
=
\rho_{\mathrm{ref}}f\boldsymbol{\Psi}^{\star}
\label{eddy_stress}
\end{equation}
is an eddy stress which vanishes
($\boldsymbol{\Psi}^{\star}=0$) at the top and bottom of the ocean.

Its physical interpretation is as a mesoscale eddy ``form drag'' and
has a status equivalent to that of the applied wind at the surface,
except that it exists in the interior of the ocean --- see the
discussion in \citet{FerreiraMarshall2006}.

The zonal ($x$) component of Eq.~(\ref{res_mtm}) is:
\begin{equation}
-fv_{\mathrm{res}}
+
\frac{1}{\rho_{\mathrm{ref}}}
\frac{\partial p}{\partial x}
=
\frac{1}{\rho_{\mathrm{ref}}}
\frac{\partial}{\partial z}
\left(\tau_x^w+\tau_x^e\right)
\simeq
\frac{1}{\rho_{\mathrm{ref}}}
\frac{\tau_x^w-\tau_x^e}{\Delta z},
\label{res_mtm_x_compt}
\end{equation}

where $\Delta z$ is the depth of the topmost layer of the ocean
model.

Note that $\tau_x^w$ is the zonal wind stress the ocean feels at the
surface, which decays with depth over the Ekman layer. In contrast,
the eddy stress, $\tau_x^e$, is zero at the surface but increases with
depth in the near-surface layers to reach magnitudes approaching the
applied wind stress in the stratified interior.

In BIG-MITgcm, the topmost layer of the model has a depth
$\Delta z=20$~m which we assume is of the same order as the Ekman
layer depth, the depth over which $\tau_x^w$ decays. Typical Southern
Ocean surface Ekman layers are a few tens of metres to about 50~m ---
and usually at or above the mixed-layer base.

We diagnose our multiple equilibrium solutions using the
framework provided by Eq.~(\ref{res_mtm_x_compt}).

\section{Complementary description of the climate states}

% ============================================================
% TABLE
% ============================================================

\begin{table}[H]
%\small
\fontsize{8.5pt}{8.5pt}\selectfont
\centering
\caption{Global annual mean values averaged over the last 30 years,
and associated standard deviations derived from interannual
variability. The acronyms in the table are: surface air temperature
(SAT), top-of-the-atmosphere budget ($R_t$), ocean surface budget
($F_s$), heat transport (HT),  northern hemisphere (NH), southern hemisphere (SH),
water-mass transport (WM), evaporation minus precipitation (E--P), sea surface temperature
(SST), sea surface salinity (SSS), and net primary productivity (NPP). The Atlantic Meridional Overturning Circulation (AMOC) intensity is computed
as the maximum between 20$^\circ$N and 30$^\circ$N and -1000 m and
-3000 m; the SAT gradient is computed by averaging from -30$^\circ$
to 30$^\circ$, and from 30$^\circ$ to 90$^\circ$, and then taking the
difference between the two.}
\label{tab:PT_global}

\begin{tabular}{lccccc}
\toprule
&
\multicolumn{1}{c}{Hot} &
\multicolumn{1}{c}{Warm} &
\multicolumn{1}{c}{Cold} &
\multicolumn{1}{c}{Waterbelt} &
\multicolumn{1}{c}{Snowball}
\\
\midrule

SAT ($^\circ$C)
& 16.64$\pm$0.06
& 13.78$\pm$0.08
& 5.0$\pm$0.1
& -14.0$\pm$0.1
& -41.11$\pm$0.02
\\

$R_t$ (W\,m$^{-2}$)
& -0.6$\pm$0.1
& -0.5$\pm$0.1
& -0.4$\pm$0.2
& -1.4$\pm$0.2
& -0.26$\pm$0.04
\\

$F_s$ (W\,m$^{-2}$)
& -0.2$\pm$0.1
& 0.0$\pm$0.2
& 0.2$\pm$0.2
& -0.1$\pm$0.2
& $ (1.9\pm 0.2)10^{-4}$
\\

HT max intensity (PW)
& 6.22 & 6.42 & 5.91 & 5.01 & 0.64
\\

HT max latitude ($^\circ$)
& 36 & 36 & 36 & 28 & 50
\\

NH sea-ice extent ($10^6$ km$^2$)
& 11.7$\pm$0.2
& 9.2$\pm$0.1
& 14.8$\pm$0.3
& 46.6$\pm$0.4
& 119$\pm$0
\\

SH sea-ice extent ($10^6$ km$^2$)
& 1.3$\pm$2
& 22.6$\pm$0.5
& 43.1$\pm$0.9
& 94.7$\pm$0.5
& 191$\pm$0
\\

$E-P$ ($10^{-8}$ kg\,m$^{-2}$\,s$^{-1}$)
& 0$\pm$1
& 0$\pm$2
& 0$\pm$1
& 0.4$\pm$0.4
& 0.03$\pm$0.01
\\

WM max intensity ($10^9$ kg\,s$^{-1}$)
& 1.22 & 1.08 & 0.94 & 0.50 & 0.00
\\

WM max latitude
& 36 & 36 & 36 & 26 & 26
\\

NH SAT gradient ($^\circ$C)
& 32.8$\pm$0.3
& 30.2$\pm$0.3
& 43.0$\pm$0.5
& 49.2$\pm$0.1
& 26.3$\pm$0.1
\\

SH SAT gradient ($^\circ$C)
& 31.3$\pm$0.1
& 39.1$\pm$0.2
& 43.6$\pm$0.3
& 47.5$\pm$0.3
& 24.4$\pm$0.1
\\

SST ($^\circ$C)
& 18.78$\pm$0.03
& 17.31$\pm$0.03
& 12.81$\pm$0.05
& 4.18$\pm$0.02
& -1.7931$\pm$0.0002
\\

SSS (psu)
& 36.001$\pm$0.009
& 36.242$\pm$0.005
& 37.625$\pm$0.009
& 38.588$\pm$0.005
& -- %36.3109$\pm$0.0002
\\

AMOC intensity (Sv)
& -1.3 & 27.2 & 43.76 & -4.76 & 13.41
\\

NH ice-sheet volume ($10^6$ km$^3$)
& 1.6 & 2.2 & 68.0 & 96.0 & 126.6
\\

SH ice-sheet volume ($10^6$ km$^3$)
& 21.2 & 21.0 & 19.0 & 24.5 & 45.4
\\

NH ice-sheet area ($10^6$ km$^2$)
& 1.3 & 1.8 & 34.8 & 65.5 & 118.0
\\

SH ice-sheet area ($10^6$ km$^2$)
& 12.7 & 12.9 & 13.6 & 20.1 & 43.8
\\

Sea-level change (m)
& 1.03 & 0 & -162.0 & -244.5 & -369.1
\\

Mean NPP (gC\,m$^{-2}$)
& 530.9 & 552.1 & 377.5 & 211.3 & 0
\\

\bottomrule
\end{tabular}
\end{table}

%%%%%%%%%%%%%%%%%%%%%%%%%%%
\begin{figure}[H]
\centering
\includegraphics[width=\textwidth]{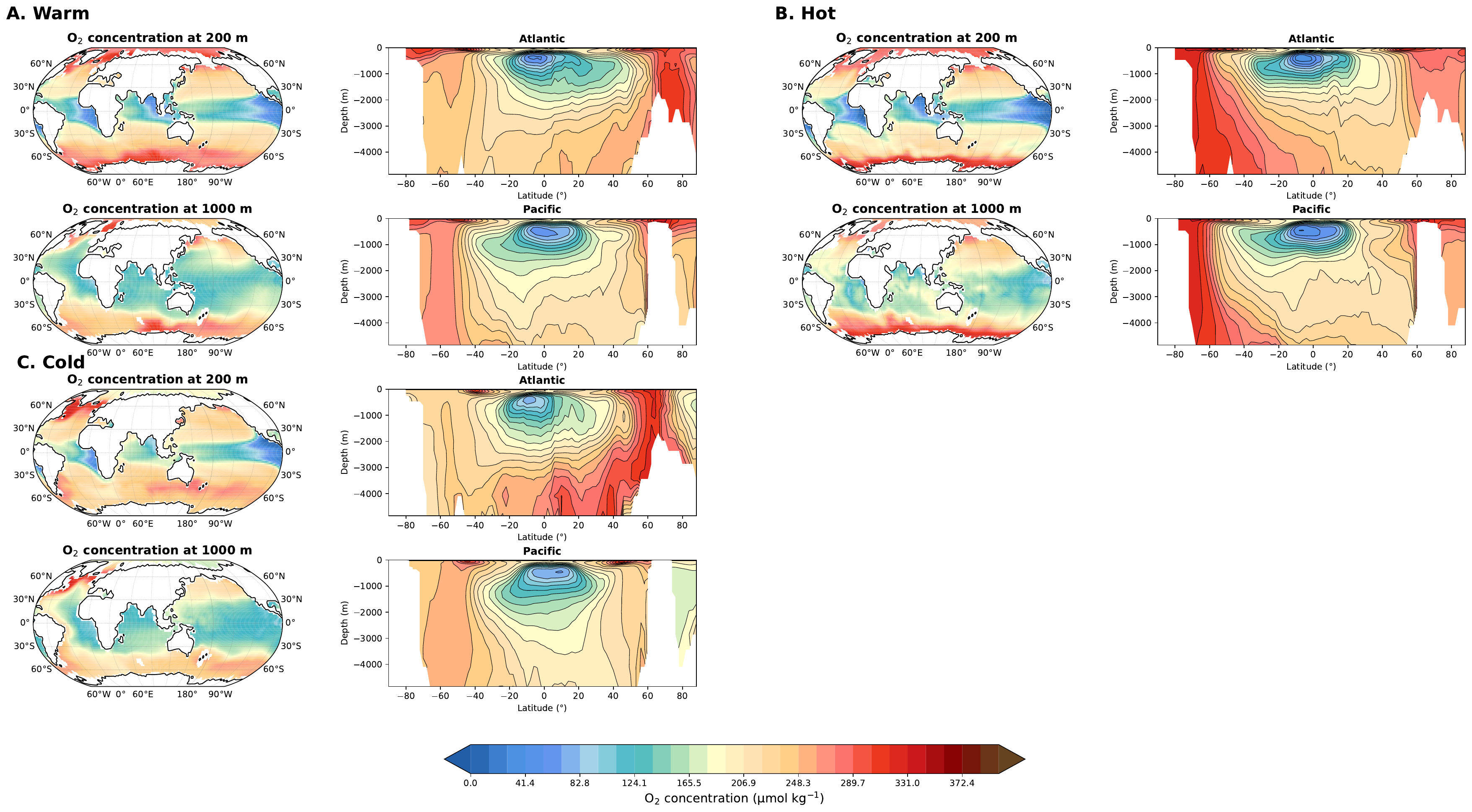}
\caption{Oxygen concentration at depths of 200 m (first map) and
1000 m (second map), together with the oxygen distributions in the
Atlantic (first subplot) and Pacific (second subplot), for the Warm
(A), Hot (B), and Cold (C) states in $\mu$mol/kg by the
blue--green--red color shading.}
\label{fig:oxygen}
\end{figure}

\begin{figure}[H]
\centering
\includegraphics[width=\textwidth]{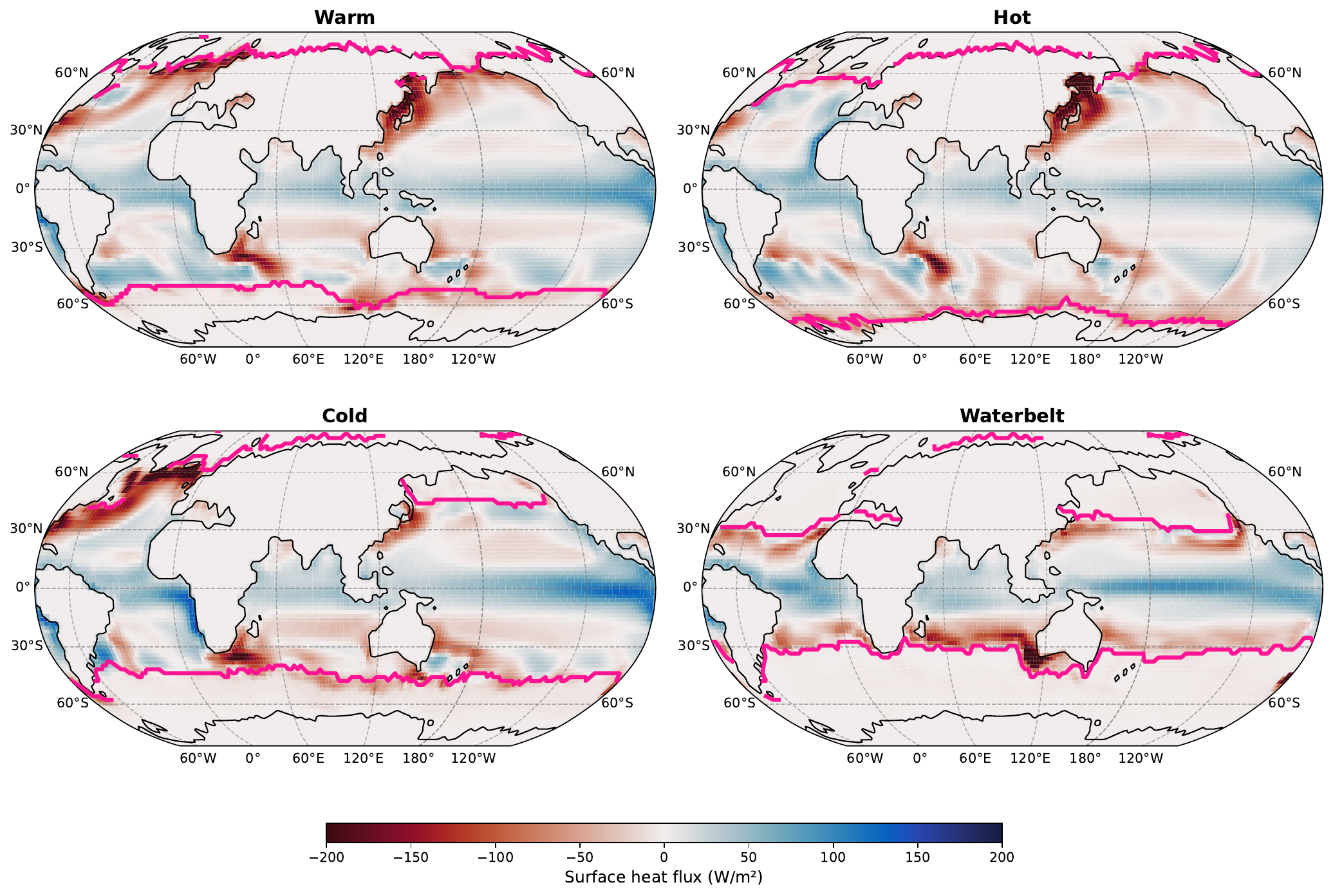}
\caption{Surface ocean heat flux for the Warm (top left), Hot (top
right), Cold (bottom left) and Waterbelt (bottom right) states.
Red shading indicates warming of the atmosphere (upward flux) and
blue shading indicates warming of the ocean (downward flux). The
annual-mean sea-ice edge is indicated by the pink line for each
state.}
\label{fig:Figure9}
\end{figure}

\begin{figure}[H]
\centering
\includegraphics[width=\textwidth]{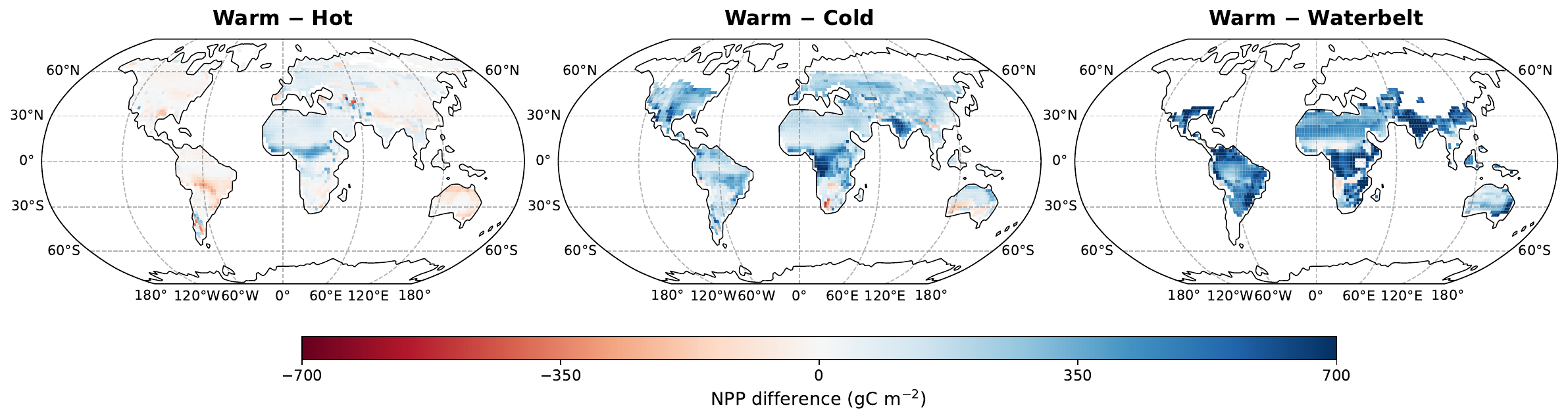}
\caption{Difference in NPP distribution relative to the Warm state
for the Hot (first panel), Cold (second panel), and Waterbelt
(third panel) states. The red--blue colour scale indicates increases
and decreases in NPP, respectively.}
\label{fig:npp_diff_map}
\end{figure}

\begin{figure}[H]
\centering
\includegraphics[width=\textwidth]{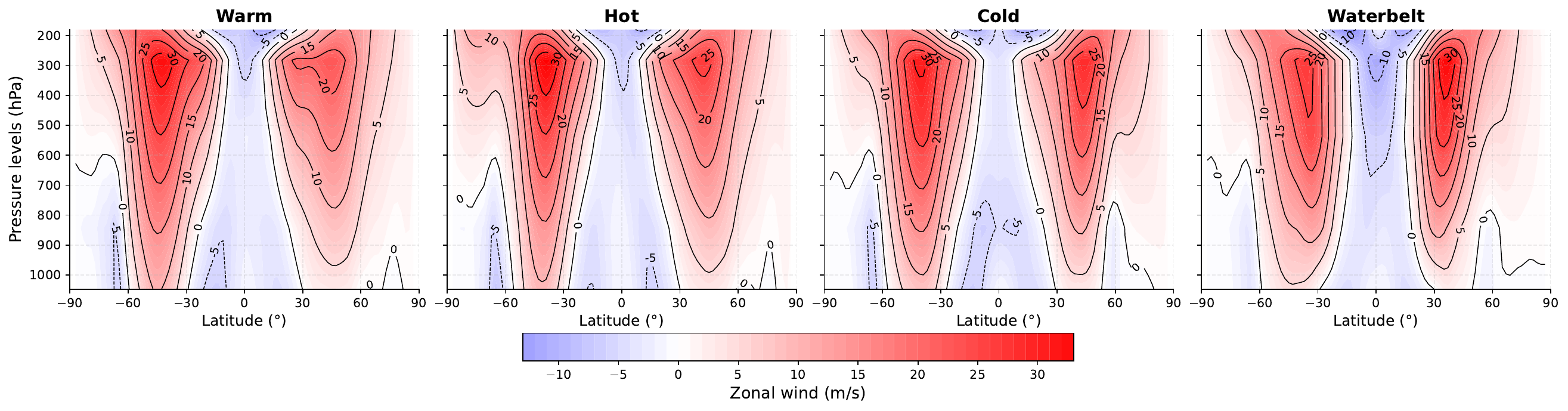}
\caption{Zonal atmospheric profiles of zonal wind for the Warm
(first panel), Hot (second panel), Cold (third panel), and Waterbelt
(fourth panel) states. The red--blue color scale indicates wind
velocity. Solid black contours represent positive values, while
dashed black contours represent negative values.}
\label{fig:zonal_wind_verticsal_4states}
\end{figure}

\begin{figure}[H]
\centering
\includegraphics[width=\textwidth]{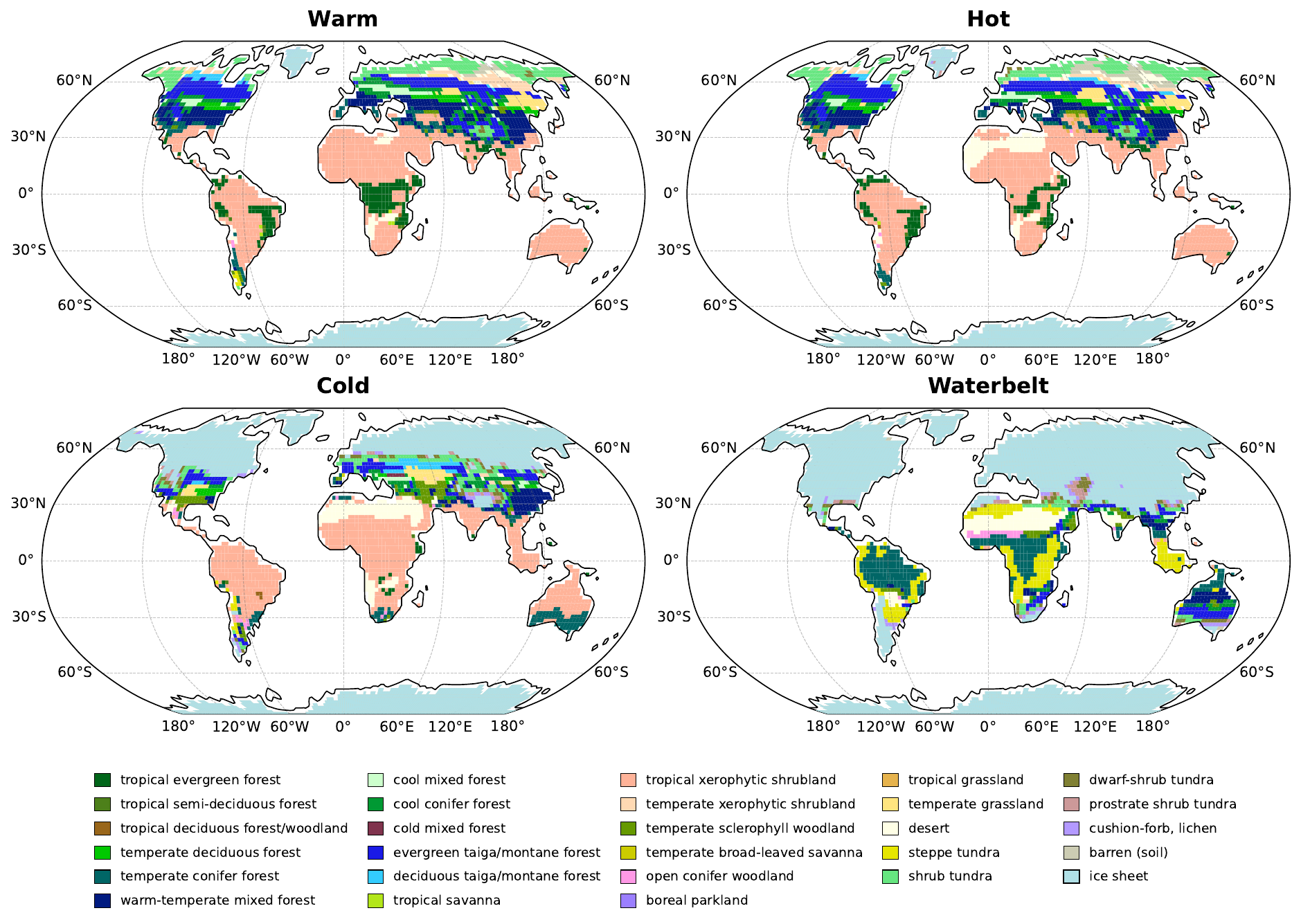}
\caption{Biome distribution for the Warm (top left), Hot (top right),
Cold (bottom left), and Waterbelt (bottom right) states. Each biome
is represented by a distinct colour, as indicated by the colour scale
above. The light blue colour represents areas covered by an ice sheet.}
\label{fig:biome_map}
\end{figure}

\begin{figure}[ht!]
\centering
\includegraphics[width=\textwidth]{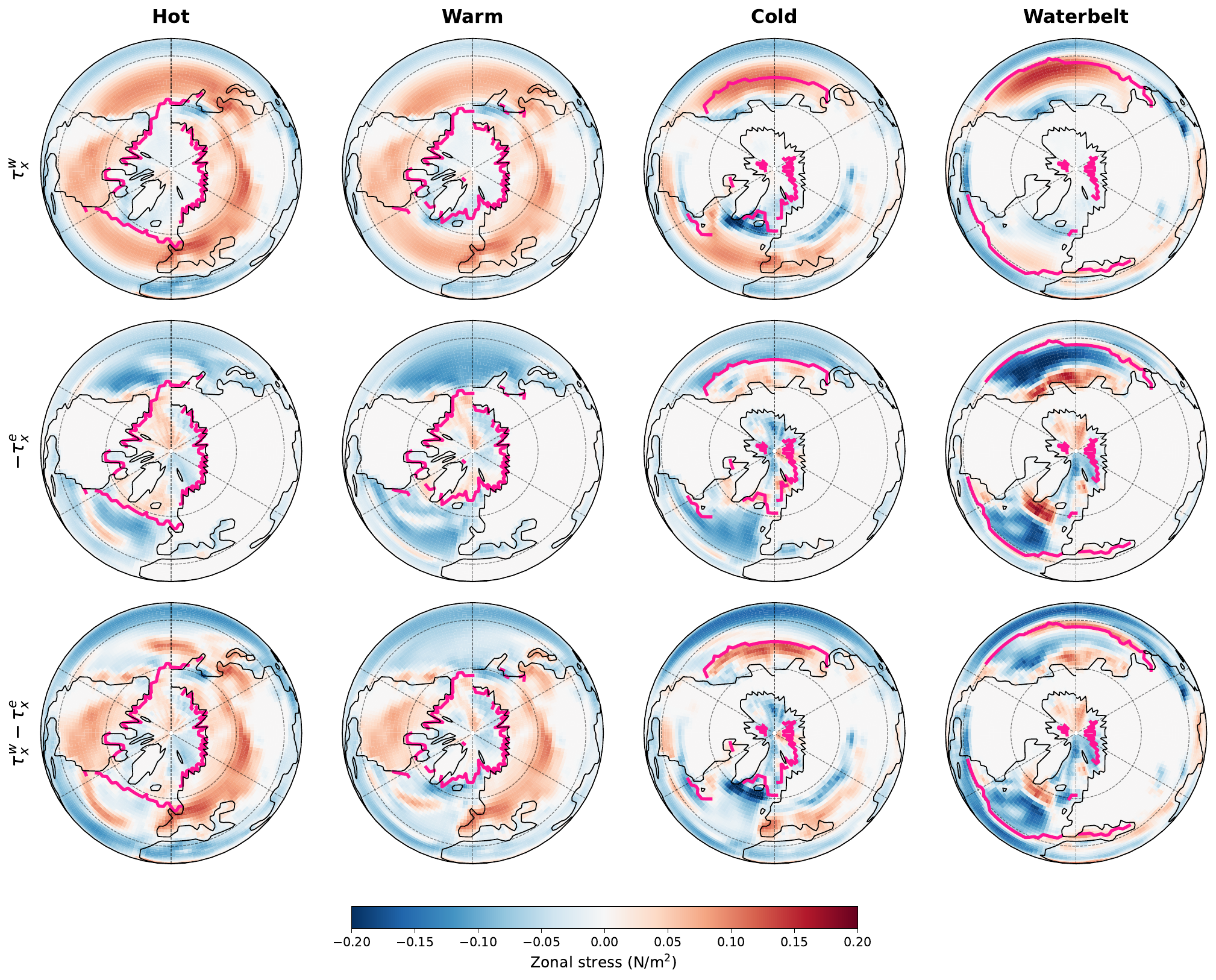}
\caption{The mean annual zonal component of the wind stress, $\tau_x^w$, of minus the eddy stress, $-\tau_x^e$, and of their sum, $\tau_x^w-\tau_x^e$, in the Northern Hemisphere, expressed in $\mathrm{N}\,\mathrm{m}^{-2}$. The pink line represents the annual mean sea-ice edge.}
\label{zonal_stress}
\end{figure}

\begin{figure}[ht!]
\centering
\includegraphics[width=0.5\textwidth]{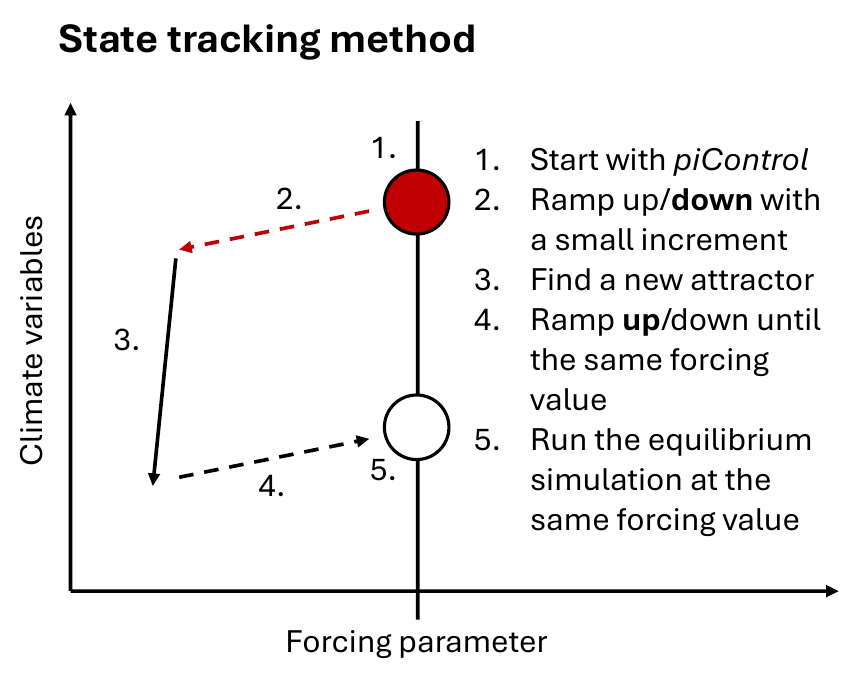}
\caption{The state tracking method is illustrated schematically. 1. Begin with a control state represented by the red circle;
2. Ramp up/down with a small increment in the forcing along the red dashed line; 3. Find a new steady state and apply the iterative procedure with \textit{BIG-MITgcm}; 4. Ramp up/down along the
black dashed line until the same initial forcing value is reached; 5. Integrate out to equilibrium at the same forcing value applying the iterative procedure with \textit{BIG-MITgcm}. }
\label{fig:state_tracking_method}
\end{figure}

\end{document}